\documentclass[nohyper]{JHEP3}
\usepackage{amsmath,amsfonts}
\usepackage[dvips]{graphicx}
\usepackage{epsfig,yfonts}
\allowdisplaybreaks[2]

\newcommand{\labell}[1]{\label{#1}}  %\mt{#1}

\newcommand{\be}{\begin{equation}}
\newcommand{\ee}{\end{equation}}
\newcommand{\bea}{\begin{eqnarray}}
\newcommand{\eea}{\end{eqnarray}}
\newcommand{\ba}{\begin{eqnarray}}
\newcommand{\ea}{\end{eqnarray}}

\newcommand{\beq}{\begin{equation}}
\newcommand{\eeq}{\end{equation}}
\newcommand{\beqa}{\begin{eqnarray}}
\newcommand{\eeqa}{\end{eqnarray}}
\newcommand{\beqar}{\begin{eqnarray*}}
\newcommand{\eeqar}{\end{eqnarray*}}

\newcommand{\reef}[1]{(\ref{#1})}

\newcommand{\eg}{{\it e.g.,}\ }
\newcommand{\ie}{{\it i.e.,}\ }

\renewcommand{\c}[1]{c_\mt{#1}}
\newcommand{\mt}[1]{\textrm{\tiny #1}}

\newcommand{\veps}{\varepsilon}
\newcommand{\X}{\mathcal{X}}
\newcommand{\Z}{\mathcal{Z}}
\newcommand{\D}{\mathcal{D}}
\newcommand{\fin}{f_\infty}

\title{Black Holes in Quasi-topological Gravity}

\author{Robert C. Myers$^1$ and Brandon Robinson$^{1,2}$\\
$^1\,$Perimeter Institute for Theoretical Physics,
Waterloo, Ontario N2L 2Y5, Canada \\
$^2\,$Department of Physics \& Astronomy and Guelph-Waterloo Physics
Institute,\\
\ \  University of Waterloo,
Waterloo, Ontario N2L 3G1, Canada \\
\\E-mail: \email{
rmyers@perimeterinstitute.ca,
 brobinson@perimeterinstitute.ca }}

\preprint{arXiv:1003.5357 [gr-qc]}

\date{\today}

\abstract{We construct a new gravitational action which includes cubic
curvature interactions and which provides a useful toy model for the
holographic study of a three parameter family of four- and higher-dimensional
CFT's. We also investigate the black hole solutions of this new gravity
theory. Further we examine the equations of motion of quasi-topological
gravity. While the full equations in a general background are
fourth-order in derivatives, we show that the linearized equations
describing gravitons propagating in the AdS vacua match precisely
the second-order equations of Einstein gravity. }

\begin{document}{\vskip 1cm}

%%%%%%%%%%%%%%%%%%%%%%%%%%%%%%%%%%%%%%%%%%%%%%%%%%%%%%%%%%%%%%
%%%%%%%%%%%%%%%%%%%%%%%%%%%%%%%%%%%%%%%%%%%%%%%%%%%%%%%%%%%%%%
%%%%%%%%%%%%%%%%%%%%%%%%%%%%%%%%%%%%%%%%%%%%%%%%%%%%%%%%%%%%%%
%%%%%%%%%%%%%%%%%%%% INTRODUCTION %%%%%%%%%%%%%%%%%%%%%%%%%%%%
%%%%%%%%%%%%%%%%%%%%%%%%%%%%%%%%%%%%%%%%%%%%%%%%%%%%%%%%%%%%%%
%%%%%%%%%%%%%%%%%%%%%%%%%%%%%%%%%%%%%%%%%%%%%%%%%%%%%%%%%%%%%%
%%%%%%%%%%%%%%%%%%%%%%%%%%%%%%%%%%%%%%%%%%%%%%%%%%%%%%%%%%%%%%

\section{Introduction}

Recently, there has been some interest in gravitational actions with
higher curvature actions in the context of the AdS/CFT correspondence.
For example, Einstein gravity in the AdS bulk defines a universality
class of CFT's in which the ratio of the shear viscosity to entropy
density is given by precisely $\eta/s=1/(4\pi)$ \cite{kss,einstein}.
However, it is understood that adding higher curvature interactions to
the bulk gravity action leads to a broader class of CFT's in which this
ratio generally depends on the value of the additional gravitational
couplings \cite{string4}. Further it is possible with some holographic
constructions to violate the famous bound conjectured by Kovtun, Son
and Starinets (KSS) \cite{kss} producing theories with
$\eta/s<1/(4\pi)$. In certain string theory constructions, the
appearance of curvature-squared interactions produces violations of the
KSS bound \cite{string} but these models only produce reliable results
in a regime where the corresponding gravitational coupling is
parametrically small. Hence at present, one can only deviate
perturbatively away from the universality class defined by Einstein
gravity in these string theory models.

However, it is also of interest to explore situations where the
gravitational couplings are finite. For example, holography can yield
new consistency conditions for the gravitational theories and their
dual CFT's. One theory which provides a useful toy model in this regard
is Gauss-Bonnet (GB) gravity. Even with a finite coupling for the
curvature-squared interaction, this theory still provides some
calculation control, which has been exploited in several recent
holographic studies \cite{EtasGB, hofman, Sin, jan3, moreunpub}.
However, GB gravity only introduces a single new coupling which limits
the range of dual CFT's which can be studied. A natural generalization
would be the further addition of interactions cubic in the curvature,
as this allows the investigation of the the full range of parameters in
the three-point function of the stress tensor \cite{qthydro}. A
straightforward extension of GB gravity would be to include the cubic
interaction of Lovelock gravity \cite{lovelock}. However, because of
the topological origin of the Lovelock terms the cubic interaction only
contributes to the equations of motion when the bulk dimension is seven
or greater. In the context of the AdS/CFT correspondence, this means
that such a term will be effective in expanding the class of dual CFT's
in six or more dimensions \cite{jan3,moreunpub}. Our key result in this
paper is to construct a new gravitational action with cubic curvature
interactions which provides a useful toy model to study a broader class
of four (and higher) dimensional CFT's, involving three independent
parameters. In the following, we describe the construction for the new
gravitational action and investigate black hole solutions in this
theory. We leave the detailed study of the properties of the dual class
of CFT's to a companion paper \cite{qthydro}.

An outline of the rest paper is as follows: We begin with a review of
black hole solutions and various aspects of these solutions in
Gauss-Bonnet (GB) gravity coupled to a negative cosmological constant
in section \ref{GBgrav}. Inspired by the GB equations of motion
determining black hole solutions, we construct a new interaction that
is cubic in curvatures and yields similar simple solutions in section
\ref{new}. Again we wish to emphasize that this interaction is not the
six-dimensional Euler density as appears in third-order Lovelock
gravity. Further, we show in appendix \ref{app-topo} that the new
interaction does not have a topological origin and hence we call the
new theory: `quasi-topological gravity.' We turn to a discussion of the
asymptotically AdS black hole solutions in section \ref{BHs}. While the
focus of this discussion is planar black holes in five dimensions, we
generalize the results to curved horizons and higher dimensions in
sections \ref{curvhor} and \ref{higherD}. In section \ref{thermo}, we
examine black hole thermodynamics in the new theory, deriving some of
the basic thermal properties of the black holes and the corresponding
plasmas in the dual CFT. We examine the equations of motion of
quasi-topological gravity in section \ref{eom}. While the full
equations in a general background are fourth-order in derivatives, we
show that the linearized equations describing gravitons propagating in
the AdS vacuum solutions are precisely the second-order equations of
Einstein gravity. We conclude with a brief discussion of our results
and future directions in section \ref{discuss}.

While we were in the final stages of preparing this paper, two related
preprints appeared in which exceptional new theories of curvature-cubed
gravity were constructed. Ref.~\cite{aninda} constructs an interesting
curvature-cubed theory in three dimensions. Up to a contribution
proportional to the six-dimensional Euler density, the curvature-cubed
interaction constructed in five dimensions by \cite{newer} is identical
to that studied here. Refs.~\cite{newer,newer2} are also able to relate
our interactions in $D\ge7$ to Weyl-invariant combinations of
curvatures combined with the six-dimensional Euler density.

\section{Black Holes in Gauss-Bonnet gravity} \label{GBgrav}

We begin here with a brief review of black holes in Gauss-Bonnet (GB)
gravity. The latter corresponds to a theory of gravity in which a
curvature-squared interaction is added with the form of the density for
the Euler characteristic of four-dimensional manifolds,
 \beq
\X_4=R_{abcd}R^{abcd}-4\,R_{ab}R^{ab}+R^2\,. \labell{GBterm}
 \eeq
Of course, this term will not affect the gravitational equations of
motion if the dimension of the spacetime is four (or lower), however,
it makes interesting contributions for $D\ge5$. GB gravity can be seen
as the simplest example of the Lovelock theories \cite{lovelock}
discussed above. As explained, despite having a higher curvature
action, the resulting equations of motion are still second-order in
(time) derivatives and this produced some interest in early discussions
of higher curvature corrections to string theory
\cite{GBheter,GBghost}. These discussions also lead to an extensive
study of black hole solutions in this theory \cite{GBbh}. More
recently, there has been some renewed interest in asymptotically AdS
black hole solutions in GB gravity \cite{GBads,GBads2}, especially in
the context of the AdS/CFT correspondence \cite{EtasGB, hofman, Sin}.
In our following, we discuss the black hole solutions focussing on GB
gravity with $D=5$ and with a negative cosmological constant:
 \beq
I = \frac{1}{16 \pi G_\mt{5}} \int \mathrm{d}^5x \, \sqrt{-g}\, \left[
\frac{12}{L^2} + R + \frac{\lambda L^2}{2}\X_4 \right]\ .
 \labell{GBact}
 \eeq
We add some comments about higher dimensions at the end of this
section.

Let us present the ansatz for the metric of five-dimensional planar AdS
black holes, which we will be using throughout the paper:
 \beq
\mathrm{d}s^2 = \frac{r^2}{L^2}\left(-N(r)^2 f(r) \,\mathrm{d}t^2
+\mathrm{d}x^2+\mathrm{d}y^2+\mathrm{d}z^2\right) +\frac{L^2}{r^2
f(r)}\, \mathrm{d}r^2\,. \labell{metric0}
 \eeq
Inserting this metric ansatz into the action \reef{GBact} (and
integrating by parts a number of times) yields
 \beq
I = \frac{1}{16\pi G_\mt{5}} \int \mathrm{d}^5x \, \frac{3 N(r)}{L^5}\,
\left[ r^4 ( 1 - f +\lambda f^2 )\right]'
 \labell{GBact2}
 \eeq
where the `prime' indicates differentiation with respect to $r$.
Schematically, the equation of motion coming from varying the lapse $N$
takes the simple form $[r^4(\cdots)]'=0$ and so the metric function $f$
is given by solving for the roots of a quadratic polynomial
\cite{GBads,GBads2}:
 \beq
\lambda f(r)^2 -f(r)+1-\frac{\omega^4}{r^4}=0\ .\labell{quadd}
 \eeq
The latter yields two solutions
 \beq
f_\pm(r) =
\frac{1}{2\lambda}\left[1\pm\sqrt{1-4\lambda\left(1-\frac{\omega^4}{r^4}\right)}\
\right]\labell{GBeom}
 \eeq
Now varying $\delta f$ yields a constraint which requires that $N= \,
constant$, which we leave unspecified for the moment. In the following,
we will consider only the solutions with $f_-$, since the other branch
with $f_+$ contains ghosts and is unstable \cite{GBghost} --- as we
will see in later sections. With the choice $f=f_-$, it is easy to
verify that the horizon appears at $r=r_h=\omega$.

In fixing the value of $N$, it is convenient to consider the solution
with $\omega=0$,
 \beq
\mathrm{d}s^2 = \frac{r^2}{L^2}\left(-N(r)^2 \fin \,\mathrm{d}t^2
+\mathrm{d}x^2+\mathrm{d}y^2+\mathrm{d}z^2\right) +\frac{L^2}{r^2
\fin}\, \mathrm{d}r^2\,. \labell{metric0a}
 \eeq
where we have adopted the notation
 \beq
\fin \equiv \lim_{r\rightarrow \infty} f(r)
=\frac{1}{2\lambda}\left[1-\sqrt{1-4\lambda}\ \right]\,.
 \labell{fin}
 \eeq
We recognize eq.~\reef{metric0a} as anti-de Sitter (AdS) space,
presented in the Poincar\'e coordinates. From $g_{rr}$ above, we also
see that the AdS curvature scale is given by $\tilde L =
L/\sqrt{\fin}$. This metric also makes apparent a convenient choice for
the lapse, namely $N^2=1/\fin$, which we adopt in the following. In the
AdS vacuum \reef{metric0a}, this ensures that any motions in the brane
directions are limited to lie within the standard light cone, \ie
$0=-\mathrm{d}t^2 +\mathrm{d}x^2+\mathrm{d}y^2+\mathrm{d}z^2$. In the
black brane solution \reef{metric0}, we still have
$\lim_{r\rightarrow\infty} N^2\,f(r)=1$ and so this comment applies in
the asymptotic region. In the context of the AdS/CFT correspondence,
the latter means that the speed of light in the boundary CFT is simply
$c=1$.

Examining the solutions in eq.~\reef{GBeom} or $\fin$ in
eq.~\reef{fin}, we see that there is an upper bound at $\lambda =
{1}/{4}$. For larger values of $\lambda$, the gravitational theory does
not have an anti-de Sitter vacuum and the interpretation of the
solutions \reef{GBeom} becomes problematic. In fact, using the AdS/CFT
correspondence to demand consistency of the dual CFT, \eg requiring
that the boundary theory is causal, imposes much more stringent
constraints on the GB coupling \cite{EtasGB,hofman}
 \beq
 -\frac{7}{36}\le\lambda\le\frac{9}{100}\ .
 \labell{causal}
 \eeq

We now turn to the thermodynamic properties of these GB black holes.
The temperature of the black brane solutions is then given by the
simple expression:
 \beq
T =\frac{1}{4\pi}\frac{r_h^2 f'|_{r_h}}{L^2}N = \frac{\omega}{\pi
L^2}N\,, \labell{hawkT}
 \eeq
which when evaluated with $N=1/\sqrt{\fin}$ becomes:
 \beq
T =  \frac{\omega}{\pi L^2} \left[\frac{1}{2}
(1+\sqrt{1-4\lambda})\right]^{1/2}\,.\labell{hawkT2}
 \eeq
The latter can be calculated by the standard technique of by
analytically continuing the metric \reef{metric0} to Euclidean time,
$\tau= -i\, t$, and choosing the periodicity of $\tau$ to ensure the
geometry is smooth at $r_h=\omega$. Next we evaluate the Euclidean
action:
 \beq
I_E[T] = \frac{1}{16\pi G_\mt{5}}\frac{V_3 \, \omega^4 N}{T L^5
\lambda}\left(\frac{r_+^4}{\omega^4}\left(12\lambda -
5+5\sqrt{1-4\lambda}\right)-4\lambda+\frac{2\lambda}{\sqrt{1-4\lambda}}\right),
 \labell{Eact3}
 \eeq
where $V_3$ is the regulator volume obtained by integrating the
$(x,y,z)$ directions.  Further we have limited the radial integration
from $r=\omega$ to $r_+$ to regulate the asymptotic or UV divergence in
$I_E$. The divergent $r_+^4$ contribution is removed with background
subtraction using the AdS vacuum \reef{metric0a}, \ie that is $I_E -
I_E^0$ remains finite in the limit $r_+ \rightarrow \infty$. Note that
in general, such a calculation would include a generalized
Gibbons-Hawking term \cite{genGibHawk}, as well as other boundary terms
to regulate the divergences in the Euclidean action \cite{counter}.
However, these surface terms do not contribute to the final result for
planar AdS black holes when we use the background subtraction approach.
Therefore we identify the free energy density as
 \beq
{\cal F} = \frac{T}{V_3}\left(I_E - I_E^0 \right)=
-\frac{(\pi\sqrt{\fin} L)^3}{16 G_\mt{5}}\, T^4\,.
 \labell{free0}
 \eeq
Then we may identify the energy and entropy densities as
 \beqa
\rho &=& -T^2\frac{d\ }{dT}\left({\cal
F}/T\right)=\frac{3(\pi\sqrt{\fin} L)^3\, T^4}{16 G_\mt{5}}
\,,\labell{rhoands0}\\
s &=& -\frac{d{\cal F}}{dT}=\frac{\left(\pi\sqrt{\fin} L T\right)^3}{4
G_\mt{5}} = \frac{1}{4G_\mt{5}}\frac{\omega^3}{L^3}\,.\nonumber
 \eeqa
One can confirm that the last result matches the entropy calculated
using Wald's techniques \cite{WaldEnt}.

These solutions are easily generalized from five to an arbitrary
spacetime dimension, $D$. In this case, the action is conveniently
parameterized as
 \beq
I = \frac{1}{16 \pi G_\mt{D}} \int \mathrm{d}^Dx \, \sqrt{-g}\, \left[
\frac{(D-1)(D-2)}{L^2} + R + \frac{\lambda\, L^2}{(D-3)(D-4)}\X_4 \right]\,.
\labell{genact0}
 \eeq
We also generalize the metric ansatz to include spherical and
hyperbolic, as well as planar, horizons:
 \beq
ds^2 = -(k +\frac{r^2}{L^2}f(r))N(r)^2 dt^2
+\frac{dr^2}{k+\frac{r^2}{L^2} f(r)} + r^2 d\ell_k^2 \labell{metric1}
 \eeq
where $d\ell^2_k$ is given by
 \beqa
k = +1: &\quad& d\Omega_{D-2}^2 \,\,\left({\rm metric\ on}\
S^{D-2}\right)\,, \nonumber\\
k =\ \ 0:  &\quad& \frac{1}{L^2}\sum_{i=1}^{D-2}\left(dx^i\right)^2\,,
\labell{genan}\\
k = -1: &\quad& d\Sigma_{D-2}^2 \,\,\left({\rm metric\ on}\
H^{D-2}\right)\,. \nonumber
 \eeqa
Note that for $k=\pm1$, the above line element has unit curvature. With
this general ansatz incorporating both curved horizons and $D$
spacetime dimensions, $f$ is determined by simply solving for the roots
of
 \beq
\lambda f(r)^2 -f(r)+1-\frac{\omega^{D-1}}{r^{D-1}}=0\,,\labell{quadd2}
 \eeq
and the solutions take the form given in eq.~\reef{GBeom} with the
replacement $\omega^4/r^4\rightarrow \omega^{D-1}/r^{D-1}$. The lapse
is again a constant and as above we choose $N=1/\sqrt{\fin}$. In
general, the horizon is determined by $f(r_h)=-k\frac{L^2}{r_h^2}$ and
so we only have $r_h=\omega$ for the planar horizons, \ie $k=0$. In the
case of the curved horizons, explicitly evaluating $r_h$ requires
solving a ($D-1$)-order polynomial in $r_h$. Hence we only have a
relatively simple solution in five dimensions where:
 \beq
r_h =\frac{1}{2}\sqrt{-2k\,L^2 + 2\sqrt{k^2 L^4 -4 k^2 \lambda L^4 +4
\omega^4}}\,.\labell{five-horz}
 \ee

\section{New Curvature-Cubed Interaction} \label{new}

As discussed in the introduction, we are motivated by considerations of
the AdS/CFT correspondence to consider a curvature-cubed theory of
gravity in five dimensions. GB gravity has a number of features which
one might want to reproduce, such as providing second-order equations
of motion and a family of exact black hole solutions. A natural
candidate to extend these properties to a curvature-cubed theory would
be the Lovelock theory where the six-dimensional Euler density is added
as a new gravitational interaction. However, this curvature-cubed
interaction would only contribute to the equations of motion in seven
and higher dimensions and hence will not contribute in the desired five
dimensions. While Lovelock's work then indicates that it should not be
possible to find an alternate action which yields second-order
equations of motion, we begin by writing the most general interaction
including all possible curvature-cubed (or more precisely,
six-derivative) interactions in five dimensions and attempt to tune the
coefficients to produce a simple equation for the black hole solutions,
as discussed for GB gravity in the previous section. We return to the
equations of motion in section \ref{eom} and we demonstrate that the
linearized equations of motion in the AdS vacuum are indeed
second-order.

Let us begin by listing a basis of the possible six-derivative
interactions:
\begin{center}
 \begin{tabular}{l l l}
 \centering
1. $R_{a\,\,b}^{\,\,c\,\,\,d}\, R_{c\,\,d}^{\,\,e\,\,\,f}\,
R_{e\,\,f}^{\,\,a\,\,\,b} \qquad$ &6. $R_a{}^{b} R_b{}^{c} R_c{}^{a}
\qquad\qquad$ &
11. $\nabla_a R_{b c} \,\nabla^a R^{b c}$\qquad \\
2. $R_{ab}^{\,\,\,\,\,\,cd}\, R_{cd}^{\,\,\,\,\,\,ef}\,
R_{ef}^{\,\,\,\,\,\, ab}$ &7. $R_a{}^{b}R_b{}^{a}R$\qquad & 12.
$\nabla^a R_{a b}\, \nabla^b R\qquad$  \\
3. $R_{a b c d}\, R^{a b c}{}_{e}\, R^{d e}$ &8. $R^3
\qquad\qquad$ & 13. $\nabla_a R \,\nabla^a R\ .\qquad$\\
4. $R_{a b c d}\,R^{a b c d}\, R \qquad$ &9.
$\nabla_a R_{b c d e} \nabla^a R^{b c d e} \qquad$ &  \\
5. $R_{a b c d}\, R^{a c}R^{b d}$ &10.
$\nabla^a \nabla^c R_{a b c d}\,R^{b d}$ &  \\
 \end{tabular}
\end{center}
In assembling this list, we have discarded any total derivatives, \eg
$\nabla^a \nabla_a \nabla^b \nabla^c R_{b c}$ and we have simplified
various expressions using the index symmetries of the Ricci and Riemann
tensors. In particular, these symmetries allow us to reduce any other
index contraction of three Riemann tensors to some combination of terms
1 and 2. Further, term 12 can be reduced to term 13 using $\nabla^a
R_{a b}= \frac{1}{2} \nabla_b R$. Similarly, using the Bianchi
identities, terms 9 and 10 can be shown to be reducible to other terms
and total derivatives as well. Hence, we are left with a list of 10
independent interactions which are cubic in curvatures. Combining all
of these interactions together in a single expression gives:
 \beqa
\sqrt{-g}\Z&=& \sqrt{-g}\left( \c1\,
R_{a\,\,b}^{\,\,c\,\,\,d} R_{c\,\,d}^{\,\,e\,\,\,f}
R_{e\,\,f}^{\,\,a\,\,\,b}+ \c2\, R_{ab}^{\,\,\,\,\,\,cd}
R_{cd}^{\,\,\,\,\,\,ef} R_{ef}^{\,\,\,\,\,\, ab} +\c3\, R_{a b c d}
R^{a b c}_{\,\,\,\,\,\,\,e} R^{d e}
 \right.\nonumber\\
&&\qquad
 +\, \c4\, R_{a b c d} R^{a b c d} R + \c5\,R_{a b c d} R^{a c}R^{b d}+\c6\, R_a^{\,\,b}
R_b^{\,\,c} R_c^{\,\,a}+ \c7\, R_a^{\,\,b} R_b^{\,\,a} R \nonumber\\
&&\left.\qquad  +\, \c8\, R^3 + \c{11}\, \nabla_a R_{bc}
\nabla^aR^{bc} + \c{13}\, \nabla_a R \nabla^a R \right)\,.\labell{ZZ}
 \eeqa

At this point, we substitute the black brane metric ansatz
\reef{metric0} and evaluate eq.~\reef{ZZ}. The next step will be to see
if the coefficients $\c{i}$ can be tuned to produce a result with the
same form as in eq.~\reef{GBeom}. In order to accomplish this task, we
integrate by parts repeatedly to put as many terms as possible in the
form $N(r)\times (\cdots)$ where the factor in brackets is independent
of the lapse function. The resulting expression then becomes
 \beqa
&\sqrt{-g}\Z &= - \frac{N}{4L^9}\left[\left(8r^9\left(\c{11}+2\c{13}\right)f^2f^{\left(6\right)}
+\left(r^8\left(236\c{11}+496\c{13}\right)f^2+r^9\left(56\c{13}
+28\c{13}\right)\right.\right.\right.\nonumber\\
&&\quad\times \left.\left.ff'\right)\right)f^{\left(5\right)}+\left(r^9\left(24\c3+48\c2
+20\c{11}+48\c8+12\c6+24\c7+12\c5+48c_{4}\right. \right.\nonumber\\
&&\quad\left.+\c{13}\right)ff''+8r^9\left(\c{11}+2\c{13}\right)
\left(f'\right)^2+r^7\left(5056\c{13} +384c_{4}+96\c6+2264\c{11}\right.\nonumber\\
&&\quad\left.+96\c2+96\c3+288\c7+960\c8+72\c5\right)f^2 +r^8\left(72\c5+288c_{4}+120\c3\right.\nonumber\\
&&\quad\left.\left.+1312\c{13}+192\c2+480\c8+192\c7+608\c{11}+84\c6
\right)ff'\right)f^{\left(4\right)} +\left(r^8\left(48c_{4}\right.\right.\nonumber\\
&&\quad\left.+12\c5+48\c8+24\c7+16\c{13}+12\c6 +8\c{11}+24\c3+48\c2\right)f\left(f'''\right)^2
\nonumber\\
&&\quad+\left(r^8\left(240\c8+96\c7+60\c3+100\c{11}+96\c2
+224\c{13}+36\c5+144c_{4}+42\c6\right)\left(f'\right)^2 \right.\nonumber\\
&&\quad+r^7\left(1494\c5+6069c_{4}+3054\c2+4380\c7+4012\c{11}
+9776\c{13}+12000\c8\right. \nonumber\\
&&\quad\left.+1794\c6+2340\c3+18\c1\right)ff'+\left(r^9\left(6\c5+24\c2
+24c_{4}+12\c7 +24\c8+12\c3\right.\right.\nonumber\\
&&\quad\left.+6\c6\right)f' +r^8\left(1296\c2+920\c{13}+384\c5+708\c3+388\c{11}+2016\c8+1536c_{4}\right. \nonumber\\
&&\quad\left.\left.+414\c6+888\c7\right)f\right)f''
+r^6\left(1584\c6+8140\c{11}+36\c1+1440\c2+1212\c5\right.\nonumber \\
&&\quad\left.\left.+4944\c7+1488\c3+5952\c4+17280\c8+19712\c{13}\right)
f^2\right)f'''-2r^9\left(\c5+4\c8\right.\nonumber\\
&&\quad\left.+\c6+2\c7+4c_{4}+2\c3+4\c2\right)\left(f''\right)^3 +\left(r^8\left(42\c3+84\c8+21\c6+84\c2+42\c7\right.\right.\nonumber\\
&&\quad\left.+21\c5+84c_{4}\right)f'+r^7\left(10896\c8+4260\c7+18\c1
+6384c_{4}+1308\c{11}+3696\c{13}
\right.\nonumber\\
&&\quad\left.\left.+1608\c5+2700\c3+4488\c2+1842\c6\right)f\right)
\left(f''\right)^2 +\left(r^6\left(26544\c{13}+306\c1\right.\right.\nonumber\\
&&\quad\left.+28092\c7+8790\c5+84000\c8+10740\c6+8916\c{11}+
17472\c2+12600\c3\right.\nonumber\\
&&\quad\left.+35088c_{4}\right)ff' + r^5\left(324\c1+5436\c5+564\c2+6960\c6+82560\c8+8676\c{11}\right.\nonumber\\
&&\quad\left.+22608\c7+6069\c3+24384c_{4}+24192\c{13}\right)f^2 +r^7\left(264\c{11}+252\c5+1056c_{4}\right.\nonumber\\
&&\quad\left.\left.+720\c7 +300\c6+624\c2+1920\c8+672\c{13}+408\c3\right)\left(f'\right)^2\right)
f''-16r^3\left(8\c3\right.\nonumber\\
&&\quad\left.+40c_{4}+400\c8+80\c7+16\c6+3\c1+ 16\c5+4\c2\right)f^3+r^6\left(434\c3
\right.\nonumber\\
&&\quad\left.+1240c_{4}+301\c5+950\c7+592\c2+2800\c8+ 9\c1+361\c6\right)\left(f'\right)^3+r^5\left(10122\c6 \right.\nonumber\\
&&\quad\left.+10236\c3+558\c1+10080\c{13}+2160\c{11}+ 29040\c7+12480\c2+8202\c5\right.\nonumber\\
&&\quad\left.+31296c_{4}+96000\right)\left(f'\right)^2 +r^4\left(-900\c{11}+5904\c3+24240\c7+504\c1
\right.\nonumber\\
&&\quad\left.\left.\left.+23520c_{4}+7248\c6+5748\c5+91200\c8 +5232\c2\right)\right)f^2f'
\vphantom{\left(r^9\right)}\right]+\cdots \labell{mgZ5}.
 \eeqa
Note that not all terms can be put in the desired form with further
integration by parts and so the `$\cdots$' indicates the presence of
spurious terms containing factors like $(N'')^2/N$, for example.
Focusing on the terms appearing explicitly in eq.~\reef{mgZ5}, we find
that choosing the values of the $\c{i}$'s as
\beqa
1.\ \ \c3 = -\frac{9}{7}\,\c1 -\frac{60}{7}\,\c2  &\qquad& 5.\ \
\c7 =  -\frac{33}{14}\,\c1 -\frac{54}{7}\,\c2
\nonumber\\
2.\ \ \c4 = \frac{3}{8}\c1+\frac{3}{2}\c2\ \ \ \ \ &\qquad &6.\ \ \c8 = \frac{15}{56}\c1+\frac{11}{14}\c2
\labell{table1}\\
3.\ \ \c5 = {\frac {15}{7}}\,\c1+{\frac {72}{7}}\,\c2\ & \qquad &7.\ \ \c{11} = 0 \nonumber\\
4.\ \ \c6= {\frac {18}{7}}\,\c1+{\frac {64}{7}}\,\c2\ &\qquad &8.\ \
\c{13} = 0 \nonumber \eeqa
reduces this expression to the following simple form
 \beq
\sqrt{-g}\Z = \frac{12}{7} \frac{N(r)}{L^9} (\c1 +2\c2)
(r^4 f^3)'\ . \labell{fiveD}
 \eeq
At the same time, the spurious terms denoted by `$\cdots$' in
eq.~\reef{mgZ5} also vanish with this choice of coefficients. It is
quite remarkable that there is enough freedom in the general action
\reef{ZZ} to produce this simple result. In fact, we are still free to
choose the (relative) values of $\c1$ and $\c2$ in constructing this
interaction. Explicitly then, if we choose $\c1=1,\ \c2 =0$, the new
curvature-cubed interaction takes the form
 \beqa
\Z_5&=& R_{a\,\,b}^{\,\,c\,\,\,d} R_{c\,\,d}^{\,\,e\,\,\,f}
R_{e\,\,f}^{\,\,a\,\,\,b} + \frac{1}{56}\left(21\,R_{a b c d}R^{a b
c d} R-72\,R_{a b c d}R^{a b c}{}_{e}R^{d e}
 \right.\nonumber\\
&&\qquad\left.+ 120\,R_{a b c d} R^{a c}R^{b d}
+144\,R_a{}^{b}R_b{}^{c}R_c{}^{a} - 132\, R_a^{\,\,b}R_b^{\,\,a}R
+15\,R^3\right)\,\labell{result5}
 \eeqa
or with $\c1=0,\ \c2 =1$,
 \beqa \Z_5'&=&
R_{ab}{}^{cd} R_{cd}{}^{ef} R_{ef}{}^{ab} +
\frac{1}{14}\left(21\,R_{a b c d}R^{a b c d} R-120\,R_{a b c d}R^{a
b c}{}_{e}R^{d e}
 \right.\nonumber\\
&&\qquad\left.+ 144\,R_{a b c d} R^{a c}R^{b d}
+128\,R_a{}^{b}R_b{}^{c}R_c{}^{a} - 108\, R_a^{\,\,b}R_b^{\,\,a}R
+11\,R^3\right)\,.\labell{result5a}
 \eeqa

The fact that we do not produce a unique interaction
%(even up to an overall constant pre-factor)
should not be surprising. Any curvature-cubed interaction can be
modified by the addition of the six-dimensional Euler density $\X_6$
without affecting the equations of motion. In fact, we can infer the
form of $\X_6$ by setting $\c1=-2\c2$, in which case eq.~\reef{fiveD}
vanishes, as it must if evaluated for the six-dimensional Euler
density. A standard normalization for the six-dimensional Euler density
is:
 \beqa
\mathcal{X}_6&=& \frac{1}{8}\,
\veps_{abcdef}\,\veps^{ghijkl}\,R_{ab}{}^{gh}\, R_{cd}{}^{ij}\,
R_{ef}{}^{kl}
 \nonumber\\
&=& 4\, R_{ab}^{\,\,\,\,\,\,cd} R_{cd}^{\,\,\,\,\,\,ef}
R_{ef}^{\,\,\,\,\,\, ab}-8\, R_{a\,\,b}^{\,\,c\,\,\,d}
R_{c\,\,d}^{\,\,e\,\,\,f} R_{e\,\,f}^{\,\,a\,\,\,b} -24\, R_{a b c
d} R^{a b c}_{\,\,\,\,\,\,\,e} R^{d e} +3\, R_{a b c d} R^{a b c d}
R
 \nonumber\\
&&\qquad\quad
 +24\,R_{a b c d} R^{a c}R^{b d}+16\, R_a^{\,\,b}
R_b^{\,\,c} R_c^{\,\,a} -12\, R_a^{\,\,b} R_b^{\,\,a} R + R^3\,,
\labell{euler6}
 \eeqa
where in the first line, $\veps_{abcdef}$ is the completely
antisymmetric tensor in six dimensions and hence the corresponding
expression only applies for $D=6$.  However, the first line also makes
clear that this expression should vanish when evaluated in five (or
lower) dimensions. This normalization corresponds to the choice $\c2=4$
and $\c1 =-8$. That is, $\X_6=4\Z_5'-8\Z_5$.

\subsection{Generalizing to $D\ge5$}\label{Dcube}

At this point, we turn to generalizing this construction to higher
dimensions. Given the freedom discussed above, we begin by setting
$\c1=1$ and $\c2=0$. Comparing to eq.~\reef{euler6} shows we are
guaranteed that, if a nontrivial interaction exists, the result will be
distinct from the six-dimensional Euler density. With this choice, we
substitute the $D$-dimensional extension of eq.~\reef{metric0} into the
action \reef{ZZ}. One then finds with a judicious choice of the
remaining $\c{i}$'s, the result can be reduced to
 \beq
\sqrt{-g}\Z_\mt{D} \sim \frac{N(r)}{L^D} \left(r^{D-1}f(r)^3\right)' \ .
\labell{hiDresult}
 \eeq
The required choice of coefficients (with $\c1=1$ and $\c2=0$) is:
\beqa
1.\ \ \c3(D)=-\frac{3(D-2)}{(2D-3)(D-4)} &\qquad &5.\ \ \c7(D)=-\frac{3(3D-4)}{2(2D-3)(D-4)}
\nonumber\\
2.\ \ \c4(D)=\frac{3(3D-8)}{8(2D-3)(D-4)}\ &\qquad &6.\ \ \c8(D)=\frac{3D}{8(2D-3)(D-4)}
\labell{table2} \\
3.\ \ \c5(D)=\frac{3D}{(2D-3)(D-4)}\ \ \,&\qquad & 7.\ \ \c{11}(D) =0
\nonumber\\
4.\ \ \c6(D)=\frac{6(D-2)}{(2D-3)(D-4)}\ \ \,&\qquad &8.\ \ \c{13}(D) =0\ .
\nonumber
\eeqa
In practice, we determined the coefficients separately for
$D=5\ldots10$ and found the general expressions above to fit the
results in all of these cases. Given these expressions, the general
form of $\Z_D$ becomes
 \beqa
\Z_\mt{D}&=& R_{a\,\,b}^{\,\,c\,\,\,d} R_{c\,\,d}^{\,\,e\,\,\,f}
R_{e\,\,f}^{\,\,a\,\,\,b} +
\frac{1}{(2D-3)(D-4)}\left(\frac{3(3D-8)}{8}R_{a b c d}R^{a b c d}
R \right.\nonumber\\
&&\quad-\,3(D-2) R_{a b c d}R^{a b c}{}_{e}R^{d e}+ 3DR_{a b c d} R^{a c}R^{b d}\labell{result}\\
&&\quad\left.+\,6(D-2)R_a{}^{b}R_b{}^{c}R_c{}^{a}-\frac{3(3D-4)}{2} R_a^{\,\,b}R_b^{\,\,a}R
+\frac{3D}{8}R^3\right)\,.\nonumber
 \eeqa
One easily verifies that this result reduces to eq.~\reef{result5} for
$D=5$.

An important note is that, with the coefficients prescribed above for
$D=6$, the resulting action \reef{hiDresult} is trivial, \ie there is
an overall factor of zero in this result. Hence $\Z_6$ does not
actually produce a nontrivial interaction which is cubic in curvatures.
One might be tempted to believe instead that $\Z_6$ yields another
topological invariant in six dimensions. Of course, there is no obvious
known invariant to which $\Z_6$ might correspond \cite{topo}. In
appendix \ref{app-topo}, we demonstrate that $\int d^6x\sqrt{g}\,\Z_6$
is not a topological invariant by explicitly evaluating this expression
for some nontrivial six-dimensional geometries. Hence, we refer to the
theory of gravity extended with our new curvature-cubed interaction as
`quasi-topological gravity.' At this point, we also note that our
construction does not yield a nontrivial curvature-cubed interaction
for $D\le4$.

Of course, one can generalize the interaction \reef{result} for $D>6$
by adding another component proportional to the six-dimensional Euler
character \reef{euler6}.  This would be equivalent to leaving $\c2$
arbitrary in our analysis above. Hence, we complete the discussion by
generalizing eq.~\reef{result5a} to higher dimensions using the
formula: $\Z_\mt{D}'=2\Z_\mt{D}+\frac{1}{4}\X_6$. The final expression
can be written as
 \beqa
\Z_\mt{D}' &=& R_{a b}{}^{c d} R_{c d}{}^{e f}R_{e f}{}^{a b}+
\frac{1}{(2D-3)(D-4)}\left(-12\left(D^2-5D+5\right)
R_{a b c d}R^{a b c}{}_{e}R^{d e}\right.\nonumber\\
&&\quad+\frac{3}{2}\left(D^2-4D+2\right) R R_{a b c d}R^{a b c d}
+12\left(D-2\right)\left(D-3\right)R_{a b c d}R^{a c}R^{b d}
\labell{result1} \\
&&\quad\left.+8\left(D-1\right)\left(D-3\right) R_{a}{}^b R_{b}{}^c
R_{c}{}^a -6\left(D-2\right)^2 R R_{a}{}^b
R_{b}{}^a+\frac{1}{2}\left(D^2 -4D +6\right)R^3\right)\,.\nonumber
 \eeqa

\section{Black Hole Solutions}\label{BHs}

We have thus far constructed a new gravitational action that includes
interactions up to cubic order in the curvature and which still yields
particularly simple equations to find black hole solutions. In this
section, we complete the study of the black holes in this new theory.
While we have written an action for the theory in arbitrary number of
spacetime dimensions, we will focus on the case $D=5$ here. Further we
begin by examining solutions of the form given in eq.~\reef{metric0}
and hence construct black holes with planar horizons
\cite{AdSCFT,GBads}. The extension of this analysis to larger $D$ and
curved horizons (with spherical or hyperbolic geometries) is
straightforward and will be discussed briefly at the end of this
section.

\subsection{Planar Black Holes} \label{planarbh}

We begin with the five-dimensional action:
 \beq
I = \frac{1}{16 \pi G_5} \int \mathrm{d}^5x \, \sqrt{-g}\, \left[
\frac{12}{L^2} + R + \frac{\lambda L^2}{2}\X_4 +\frac{7\mu L^4}{4}\Z_5
\right]
 \labell{act51}
 \eeq
which extends the GB action with the addition of the curvature-cubed
interaction $\Z_5$. Next, as in section \ref{GBgrav}, we consider the
following metric ansatz:
 \beq
\mathrm{d}s^2 = \frac{r^2}{L^2}\left(-N(r)^2 f(r) \,\mathrm{d}t^2
+\mathrm{d}x^2+\mathrm{d}y^2+\mathrm{d}z^2\right) +\frac{L^2}{r^2
f(r)}\, \mathrm{d}r^2\,. \labell{metric0b}
 \eeq
Evaluating the action \reef{act51} with this metric then yields
 \beq
I=\frac{1}{16\pi G_5} \int \mathrm{d}^5x \, \frac{3 N(r)}{L^5}\, \left[
r^4 ( 1 - f +\lambda f^2 +\mu f^3)\right]'
 \labell{act52}
 \eeq
where the prime again denotes a derivative with respect to $r$. The
variation $\delta N$ now yields
 \beqa
 \left[ r^4(1-f+\lambda f^2 + \mu f^3)
\right]' &=& 0\nonumber\\
\qquad \Longrightarrow \qquad 1-f+\lambda f^2 +\mu f^3 &=&
\frac{\omega^4}{r^4}\,. \labell{constr1}
 \eeqa
Similarly, satisfying the equation produced by taking the variation of
$f$ requires that either $N'=0$ or $-1+2\lambda f+3\mu f^2=0$. Since
the latter is generally inconsistent with eq.~\reef{constr1}, we arrive
at $N = constant$. As in section \ref{GBgrav}, we choose $N^2=1/\fin$
where $\fin \equiv \lim_{r\rightarrow \infty} f(r)$. This choice
ensures that the speed of light in the boundary metric is just one.

We are now left with a cubic equation \reef{constr1} to solve for
$f(r)$.  To do so, we first make the substitution $f = x -
\frac{\lambda}{3\mu}$, with which eq.~\reef{constr1} becomes:
 \beq
x^3-3\left(\frac{ 3\mu +\lambda^2}{9\mu^2}\right)x +2\left(
\frac{2\lambda^3+9\lambda\mu +27\mu^2(1-\frac{\omega^4}{r^4})}{54\mu^3}
\right)=0\,.  \labell{cube1}
 \eeq
This expression is further simplified by defining
 \beq
p = \frac{3\mu +\lambda^2}{9\mu^2} \qquad\qquad q= -\frac{2\lambda^3 +
9\mu \lambda +27\mu^2 (1 -\frac{\omega^4}{r^4})}{54\mu^3}\,.
\labell{pnq}
 \eeq
We then arrive at the depressed form of the equation:
 \beq
x^3 - 3p\,x -2q=0\,.\labell{cube2}
 \eeq
In the following discussion, note that the $r$-dependence is entirely
contained in the coefficient $q$. Before proceeding, we observe that
there are three distinct cases depending on the sign of the
discriminant, $\D= q^2 -p^3$, of eq.~\reef{cube2} with the following
results:
\begin{enumerate}
\item $q^2 -p^3 >0 \Rightarrow\,$ 1 real root and 2 complex roots
    conjugate to one another
\item $q^2 - p^3 <0 \Rightarrow\, $ 3 unequal real roots
\item $q^2-p^3 =0 \Rightarrow\, $ 3 real roots, at least 2 of which
    must be equal
\end{enumerate}

Assuming that $p\neq 0$, we define
 \beqa
\alpha &=& \left(q +\sqrt{q^2 -p^3} \right)^{\frac{1}{3}}\,,
 \labell{albet}\\
\beta &=& \left(q - \sqrt{q^2-p^3} \right)^{\frac{1}{3}}\,,
 \nonumber
 \eeqa
which allows the roots of eq.~\reef{cube2} to be written in the simple
form using Cardano's formula. Shifting these roots as above yields the
following solutions:
 \beqa
f_1 &=& \alpha +\beta-\frac{\lambda}{3\mu}\,,
 \nonumber\\
f_2 &=& -\frac{1}{2}(\alpha + \beta) +i \frac{\sqrt{3}}{2}
(\alpha - \beta)-\frac{\lambda}{3\mu}\,, \labell{roots1}\\
f_3 &=& -\frac{1}{2}(\alpha + \beta) -i \frac{\sqrt{3}}{2} (\alpha
-\beta)-\frac{\lambda}{3\mu}\,.\nonumber
 \eeqa
If $\D=q^2 - p^3 >0$, $\alpha$ and $\beta$ can be taken as real and
$x_1$ corresponds to the single real root. In this regime, the solution
is then given by $f=x_1- \frac{\lambda}{3\mu}$. If $\D=q^2 - p^3 <0$,
$\alpha$ and $\beta$ are necessarily complex but implicitly
eq.~\reef{roots1} still yields three unequal real roots. In this case,
$\alpha$ and $\beta$ can be chosen to have conjugate phases, \ie
 \beq
 \alpha=\sqrt{p}\ e^{i\theta/3}\quad{\rm and}\quad
 \beta=\sqrt{p}\ e^{-i\theta/3} \labell{conjugate}
 \eeq
where $\cos{\theta}={q}/{p^{\frac{3}{2}}}$ and
$\sin{\theta}=\sqrt{p^3-q^2}/{p^{\frac{3}{2}}}$. Here, we are using the
fact that $p$ is always positive in this domain. The solutions may then
be cast in the explicitly real but implicit form:
 \beqa
f_1 &=& 2\sqrt{p}\cos{\frac{\theta}{3}}-\frac{\lambda}{3\mu}\,,\nonumber\\
f_2 &=&
-\sqrt{p}\left(\cos{\frac{\theta}{3}}+\sqrt{3}\sin{\frac{\theta}{3}}\right)
-\frac{\lambda}{3\mu}
\,,\labell{roots2} \\
f_3 &=&-\sqrt{p}\left(\cos{\frac{\theta}{3}}
-\sqrt{3}\sin{\frac{\theta}{3}}\right)
-\frac{\lambda}{3\mu}\,.\nonumber
 \eeqa
While the precise form of $f(r)$ is determined by eqs.~\reef{roots1}
and \reef{roots2}, these results offer little insight into the physical
properties of the corresponding solutions, \eg which solutions actually
correspond to black holes. However, we will see below that much of the
physics can be inferred directly from the cubic equation
\reef{constr1}.

At this point, it is convenient to consider the AdS vacuum solutions.
As discussed in section \ref{GBgrav}, the latter can be found by
setting $f(r)$ to be constant (\ie setting $\omega=0$) or alternatively
taking the limit $r\rightarrow\infty$. Setting $\omega=0$ and
$f(r)=\fin$ in eq.~\reef{constr1} yields:
\beq h(\fin)\equiv
1-\fin+\lambda\fin^2+\mu\fin^3=0\ . \labell{vacsol}
\eeq
With the choice $N^2=1/\fin$, the five-dimensional metric
\reef{metric0} becomes
\beq ds^2 = \frac{r^2}{L^2}\left(-dt^2+dx^2+dy^2+dz^2
\right)+\frac{L^2}{\fin} \frac{dr^2}{r^2}\,, \labell{vacmet} \eeq
which corresponds to the metric for $AdS_\mt{5}$ in Poincar\'e
coordinates. Further from the $g_{rr}$ component, we see that the
radius of curvature of the $AdS_\mt{5}$ spacetime is
\beq \tilde{L}^2=L^2/\fin\ . \labell{vaccurv} \eeq
Implicitly, we have assumed $\fin>0$, which is not always the case ---
see discussion below.

Now let us examine these solutions as functions of the  couplings, \ie
in the $\mu-\lambda$ plane. In particular, if we insert the asymptotic
limit $r\rightarrow\infty$, the discriminant of eq.~\reef{cube2} is
useful in determining the number of vacuum solutions at different
points in the parameter space. As seen in eq.~\reef{pnq}, $p$ is
unchanged in this limit since it is independent of $r$ but $q$ is
slightly simplified:
 \beq
p = \frac{3\mu +\lambda^2}{9\mu^2} \qquad\qquad q_\infty=
-\frac{2\lambda^3 + 9\mu \lambda +27\mu^2 }{54\mu^3}\,. \labell{pnq2}
 \eeq
Given the previous discussion, we see that eq.~\reef{vacsol} yields 3
real solutions for $\D_\infty=q_\infty^2-p^3\le0$ and 1 real solution
for $\D_\infty>0$. The vanishing of the discriminant, $\D_\infty =0$,
reduces to a quadratic equation for $\mu$ with solutions
 \beq
\mu ={\frac {2}{27}}-\frac{\lambda}{3} \pm {\frac {2}{27}}\,\left(
1-3\lambda\right)^{3/2}\,. \labell{D0curves}
 \eeq
Eq.~\reef{D0curves} generates the two (upper) curves in the
$\mu$-$\lambda$ plane shown in figures~\ref{fig:Phase} and
\ref{fig:Phase2}. In the region bounded by these two curves,
$\D_\infty<0$ and there are three vacuum solutions while outside of
these two curves, $\D_\infty>0$ with one vacuum solution --- as long as
$\mu\ne0$. Of course, $\mu=0$ is a special axis in the parameter space
corresponding to the Gauss-Bonnet theory discussed in section
\ref{GBgrav}. Note that the positive branch of eq.~\reef{D0curves}
crosses the $\lambda$-axis at $\lambda = 1/4$, which is precisely the
critical coupling in the GB gravity. Recall that in GB gravity for
$\lambda<\frac{1}{4}$, there are two vacua while no vacuum solutions
exist for $\lambda>\frac{1}{4}$ .
\FIGURE[ht]{
\includegraphics[width=0.8\textwidth]{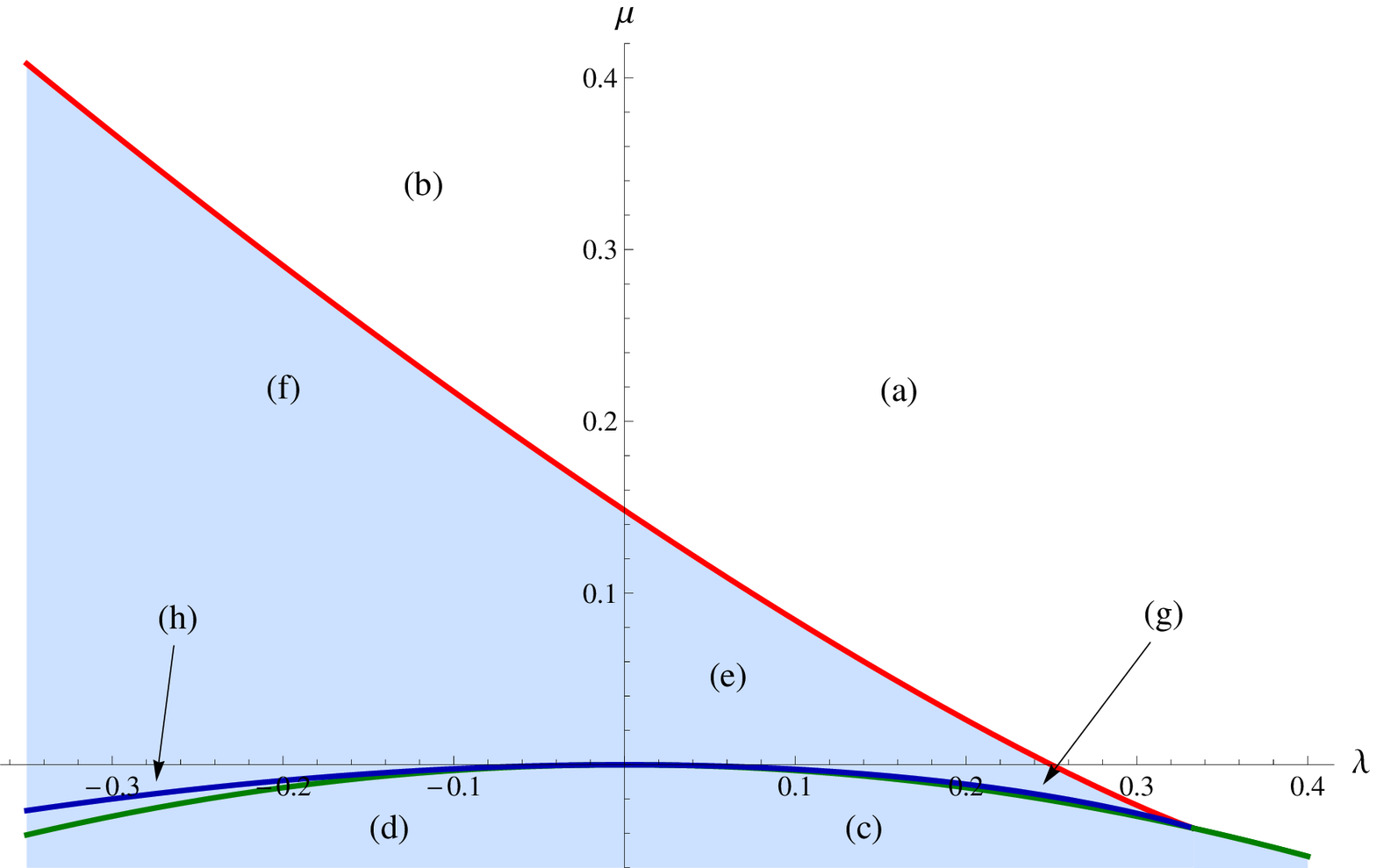}
\caption{The red and blue curves indicate the positive and negative
branches of eq.~\reef{D0curves}, respectively, where $\D=0$. The region
bounded by these two curves is where $\mathcal{D}<0$. The green curve
indicates $p=0$. The letter labels refer to the various regions
described in table \ref{table0}. The blue shaded region indicates those
couplings for which there exist asymptotically AdS black
holes.}\label{fig:Phase}}
\FIGURE[ht]{
\includegraphics[width=0.8\textwidth]{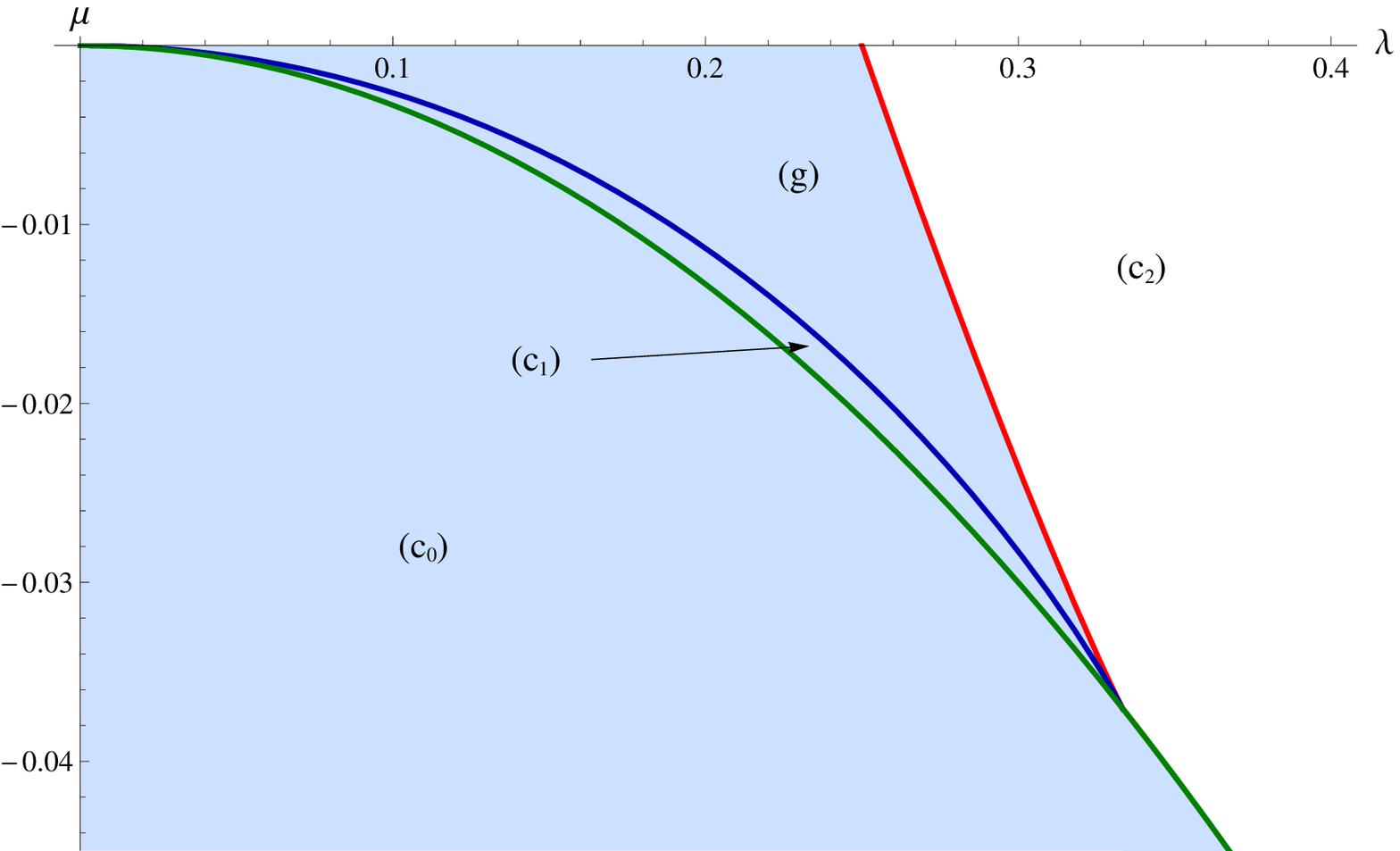}
\caption{A closer examination of the lower right quadrant of figure
\ref{fig:Phase}, covered by the regions denoted (c) and (g) in table
\ref{table0}. All three curves intersect at $(\lambda,\mu)
=(\frac{1}{3}, -\frac{1}{27})$, where the two branches of
eq.~\reef{D0curves} end. Region (g) is bounded by these two branches.
We divide (c) into: (c$_0$) where $p<0$, (c$_1$) and (c$_2$) with $p>0$
and to the left and right of (g), respectively.}\label{fig:Phase2}}

Below we will find that another interesting boundary in the
$\mu$-$\lambda$ plane is $p=0$, which corresponds to the lowermost
(green) curve in figures~\ref{fig:Phase} and \ref{fig:Phase2} -- note
that $p=0$ is always below both branches of $\D=0$, except at the
points $(\lambda,\mu)=(0,0)$ and $(\frac{1}{3}, -\frac{1}{27})$. From
eq.~\reef{pnq2}, we see that $p=0$ simply corresponds to $\mu =
-\frac{\lambda^2}{3}$. A distinguishing feature of $p=0$ is that
eq.~\reef{cube1} becomes a perfect cubic equation.

As a point of clarification, we should note that when eq.~\reef{vacsol}
yields real roots, the value of $\fin$ may be either positive (as
assumed above) or negative. For example, it is easy to see that if
$\mu>0$ one of the roots must be negative (since
$h(\fin\rightarrow-\infty)\simeq \mu\fin^3<0$ while $h(\fin=0)=1>0$).
Consistency demands $N^2>0$ and so we can not use the same choice for
the lapse when $\fin<0$. If instead we choose $N^2=1/|\fin|$, the
metric \reef{metric0} becomes
\beq ds^2 = \frac{r^2}{L^2}\left(dt^2+dx^2+dy^2+dz^2
\right)-\frac{L^2}{|\fin|} \frac{dr^2}{r^2}\,, \labell{vacmet2} \eeq
which corresponds to a particular set of coordinates on
five-dimensional de Sitter space, where we observe that $r$ plays the
role of time. The radius of curvature of this $dS_\mt{5}$ spacetime is
$\tilde{L}^2=L^2/|\fin|$. Even though with $\fin<0$, eqs.~\reef{roots1}
or \reef{roots2} still yield nontrivial solutions $f(r)$, however, the
corresponding metrics should be interpreted as (singular) cosmological
solutions rather than black holes. Hence we will not consider solutions
with $\fin<0$ further here.

We summarize our results in table \ref{table0}, which enumerates the
various kinds of vacuum solutions in different regimes of the parameter
space. Note that in this table, we have categorized the AdS vacua as
either stable or `ghosty,' \ie whether or not the graviton is a ghost
in a particular AdS vacuum. Recall that for GB gravity, we discarded
one branch of the solutions in eq.~\reef{GBeom} because the analysis of
\cite{GBghost} showed that the graviton was a ghost in these
backgrounds. A similar analysis applies for quasi-topological gravity,
as we will see in section \ref{eom}. The key factor distinguishing this
feature of the various vacua is the slope of the cubic equation
determining $\fin$:
 \beq
h'(\fin)\equiv -1+2\lambda\fin+3\mu\fin^2\,.
 \labell{vacslope}
 \eeq
In section \ref{eom}, we show that this expression appears as a
pre-factor in the kinetic term for gravitons propagating in a given AdS
vacuum. The kinetic term has the usual sign when $h'(\fin)<0$ and the
wrong sign when $h'(\fin)>0$. Hence the stable or ghost-free AdS vacua
in table \ref{table0} are distinguished by having $h'(\fin)>0$. Given
that this factor is simply the slope of $h(\fin)$, it is easy to see
that since $h(\fin=0)=1$ then if there is one AdS vacuum (\ie one root
with $\fin>0$), it will be ghost-free. Similarly if there is more than
one AdS vacuum, one of these will contain ghosts.
\begin{table}
\begin{center}
\begin{tabular}{|c|c|c|c|c|c|c|c|}
\hline
\ &$\D_\infty$&$\mu$&$\lambda$&Stable AdS&Ghosty AdS& dS& BH solution\\
\hline
a& +& +& +& 0 & 0 & 1 & -\\
b& +& +& --&  0 & 0 & 1 & -\\
c& +& --& +& 1 & 0 & 0 & $f_1$\ \ (in c$_0$,c$_1$) \\
d& +& --& --& 1 & 0 & 0 & $f_1$ \\
e& --& +& + & 1 & 1 & 1 & $f_3$\\
f& --& +& --& 1 & 1 & 1 & $f_3$\\
g& --& --& +& 2 & 1 & 0 & $f_2$\\
h& --& --& --& 1 & 0 & 2 & $f_1$ \\
\hline
\end{tabular}
\caption{Table of various vacua and black hole solutions. The column
labeled `BH solution' indicates which root \reef{roots1} yields the
nonsingular black hole solution. In case (c), the black hole solution
is only realized in the regions denoted (c$_0$)  and (c$_1$) in figure
\ref{fig:Phase2}.} \label{table0}
\end{center}
\end{table}

To gain some further insight into the black hole solutions, we return
to eq.~\reef{constr1} which we re-write as
 \beq
\tilde{h}(f)\equiv\left(1-\frac{\omega^4}{r^4}\right)-f+\lambda f^2
+\mu f^3 =0 \,.
 \labell{rewrite}
 \eeq
Now in the asymptotic limit $r\rightarrow\infty$, we simply recover
eq.~\reef{vacsol} and the roots match the vacuum solutions $\fin$. We
can regard the effect of $r$ decreasing through finite values as
reducing the `constant' term in the cubic polynomial of $f$ and as a
result, the roots of $\tilde{h}(f)$ shift away from $\fin$.

To illustrate various possibilities, figure \ref{fig:hf1vac} plots an
example of $\tilde{h}(f)$ in case (g) with three AdS vacua -- see table
\ref{table0}. First we consider the smallest root $f_2$, which
asymptotically reaches the AdS vacuum solution with the smallest value
of $\fin$. As shown when $r$ decreases, this root decreases moving
monotonically to the left until it reaches $f_2=0$ at $r=\omega$. As
$r$ shrinks to even smaller values, $f$ becomes negative and the
solution becomes singular with $f_2\rightarrow-\infty$ as
$r\rightarrow0$. Of course, this behaviour is precisely that of a black
hole with a horizon at $r=\omega$.
\FIGURE[ht]{
\includegraphics[width=0.8\textwidth]{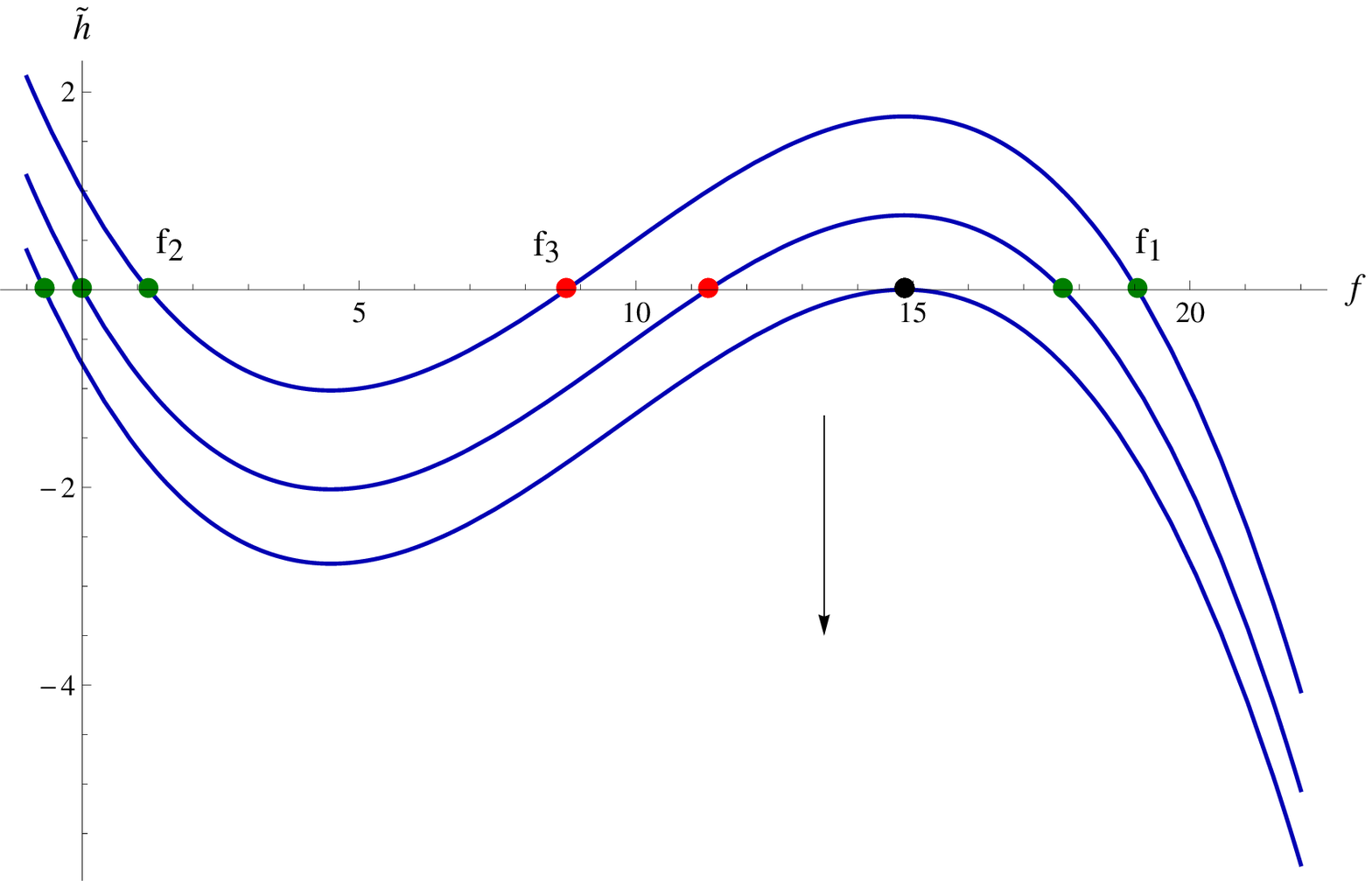}
\caption{Graph of $\tilde{h}(\fin)$ for $\mu =-0.005$ and
$\lambda=0.145$. The curve shifts down as $r$ decreases. Here, the
three curves correspond to $\omega^4/r^4=0$, 1 and 1.752 from the top
to bottom. The slope is negative for the roots, $f_1$ and $f_2$,
indicating that these are stable solutions while it is negative for
$f_3$ indicating the graviton is a ghost in this background. At
$\omega^4/r^4=1.752$, the roots $f_1$ and $f_3$ coalesce and a
curvature singularity appears in both solutions.}\label{fig:hf1vac}}

Next consider the second root $f_3$ in figure \ref{fig:hf1vac}. In this
case, the slope $\tilde{h}'(f)$ is positive and so the corresponding
AdS vacuum contains ghosts. Note that as $r$ decreases (or
$\omega^4/r^4$ increases), the root now moves to the right, \ie $f_3$
grows as we move to the interior of the solution. This behaviour is
problematic as it corresponds to a negative mass solution and it seems
to be connected to the ghost problems. As we discuss below, the
solution reaches a naked singularity at $r=r_0$ ($=1.150\omega$, in
this particular example) where the two roots, $f_1$ and $f_3$,
coalesce. One could overcome the problem with negative masses by simply
choosing the integration constant $\omega^4<0$. In this case, $f_3$
moves to the left with decreasing $r$ but a naked singularity is still
produced when the roots $f_2$ and $f_3$ coalesce.

Finally we turn to $f_1$, the largest of the three roots in figure
\ref{fig:hf1vac}. Here again, the root moves to the left as $r$
decreases so that $f_1$ decreases as we move to the interior of the
geometry indicating a positive mass. However, as noted above, this root
coalesces with $f_3$ at $r=r_0$ and becomes complex for smaller values
of $r$. Defining $f_1(r_0)=f_0$ and Taylor expanding eq.~\reef{rewrite}
about this point, we find
 \beq
f_1(r)\simeq f_0+\frac{2}{\gamma}\,\frac{\omega^2}{r_0^2}\,\left(
\frac{r-r_0}{r_0}\right)^{1/2}\quad{\rm where}\
\gamma^2=-\frac{1}{2}\tilde{h}''(f_0)=\frac{3}{2}|\mu| f_0-\lambda\,.
 \labell{expand9}
 \eeq
Further calculating the curvature using this result yields
 \beq
R_{abcd}R^{abcd}\propto\frac{\omega^4}{r_0L^4}\, \frac{1}{(r-r_0)^3}
 \labell{expand99}
 \eeq
showing that the spacetime has a naked singularity at this point. One
could again examine these solutions with $\omega^4<0$. However, in this
case, $f_1$ moves to the right, indicating a negative mass, and a naked
singularity arises with $f_1\rightarrow+\infty$ as $r\rightarrow0$.

This discussion shows that in case (g) from table \ref{table0}, only
the solution $f_2$ corresponds to an asymptotically AdS black hole. The
other roots, $f_1$ and $f_3$, are both asymptotically AdS but produce
spacetimes with naked singularities. Examining the other cases in the
table in a similar way, one finds that in each parameter regime with an
AdS vacuum, there is a single black hole solution corresponding to the
smallest positive root of eq.~\reef{rewrite}. The only exception is
case (c) where we must also be in the regions denoted (c$_0$) or
(c$_1$). These restrictions are related to the possibility that a naked
singularity will arise if the function $\tilde{h}(f)$ is not monotonic
in the range $f\in [0,\fin]$, as explained for the root $f_1$ in the
example above -- see also figure \ref{fig:hf0}. First of all, in the
region (c$_0$), $p$ is negative and $\tilde{h}(f)$ has no extrema at
all. In regions (c$_1$) and (c$_2$), $p>0$ and so one must examine the
extrema $f_0$ of $\tilde{h}(f)$. In region (c$_1$) to the left of (g)
where $\D_\infty<0$, both of the extrema $f_0> \fin$ and so
$\tilde{h}(f)$ is monotonic in the desired range. On the other hand,
one finds $0<f<\fin$ in region (c$_2$) to the right of (g). Therefore
in this parameter regime, the solution develops a naked singularity
when $r$ reaches the value where $f=f_0$. However, the solution
corresponds to a black hole with smooth event horizon for parameters in
the regions (c$_0$) and (c$_1$). All of our results with regards to
which root yields a black hole solution are summarized in table
\ref{table0}.
\FIGURE[ht]{
\includegraphics[width=0.8\textwidth]{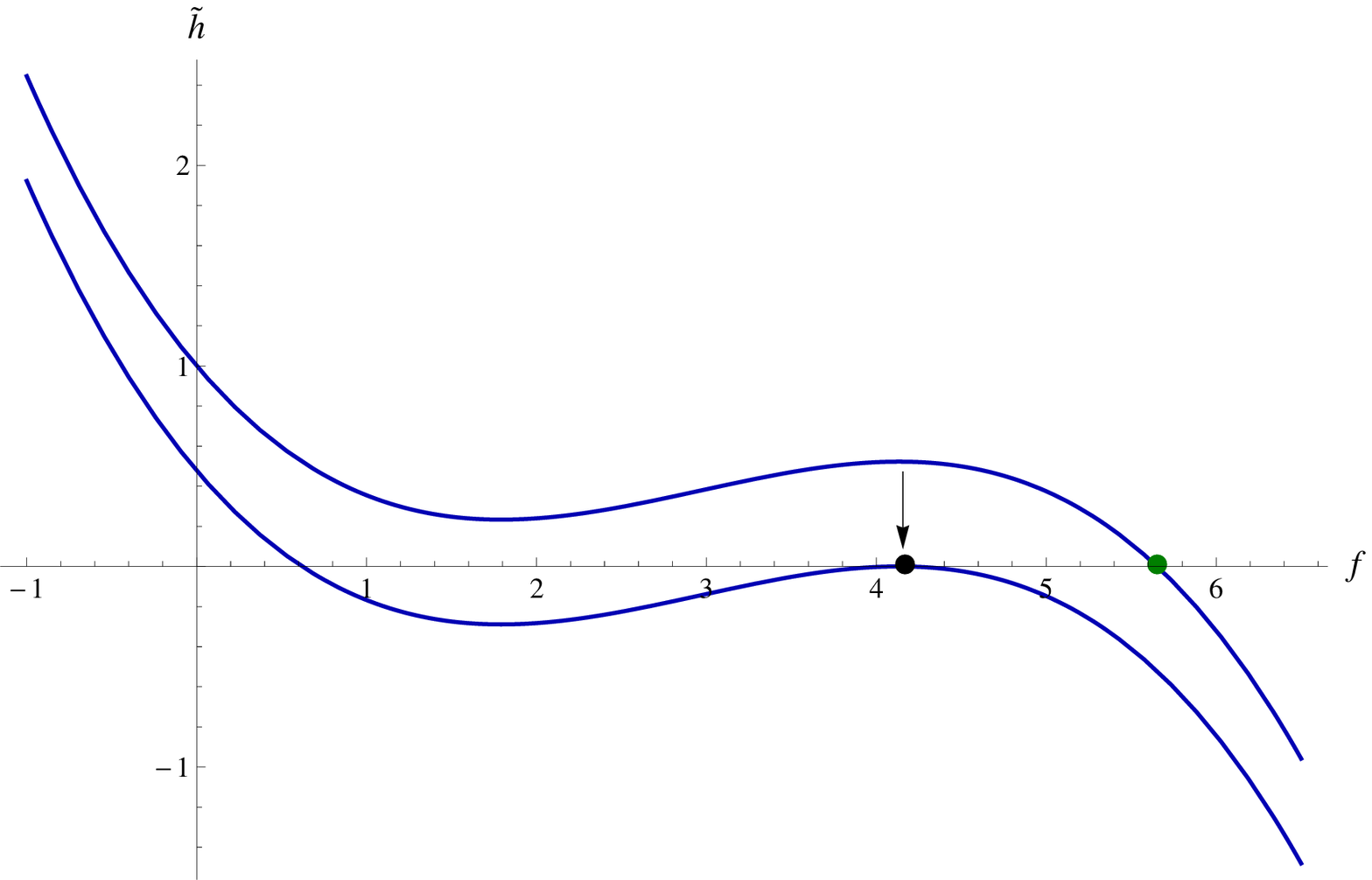}
\caption{The function $\tilde{h}(f)$ plotted for $\mu= -0.045$ and
$\lambda = 0.4$, typical parameters in the region (c$_2$). The single
(real) root $f_1$ corresponds to a ghost-free asymptotically AdS
solution. However, when the radius reaches a value where
$\tilde{h}'(f)=0$ at the root, indicated with the black dot, the
geometry becomes singular.} \label{fig:hf0}}

\subsection{Curved horizons} \label{curvhor}

As in eq.~\reef{metric1}, we can again generalize the metric ansatz to
include spherical and hyperbolic, as well as planar, horizons:
 \beq
ds^2 = -\left(k +\frac{r^2}{L^2}f(r)\right) N(r)^2 dt^2
+\frac{dr^2}{k+\frac{r^2}{L^2} f(r)} + r^2 d\ell_k^2 \labell{metric12}
 \eeq
where $d\ell_k^2$ is given by
 \beqa
k = +1: &\quad& d\Omega_{3}^2\,, \nonumber\\
k =\ \ 0:  &\quad& \frac{1}{L^2}\left(dx^2+dy^2+dz^2\right)\,,
\labell{genan2}\\
k = -1: &\quad& d\Sigma_{3}^2 \,. \nonumber
 \eeqa
As in eq.~\reef{genan}, we have above the metric on a unit three-sphere
for $k=+1$ and on a three-dimensional hyperbolic plane with unit
curvature for $k=-1$. The analysis at the beginning of section
\ref{planarbh} follows through unchanged. Implicitly, we will assume
$N(r)^2=1/\fin$ but more importantly, the solutions are again
determined by eq.~\reef{constr1}:
 \beq
1-f+\lambda f^2 +\mu f^3 = \frac{\omega^4}{r^4}\,. \labell{constr111}
 \eeq
Hence one arrives at the same solutions for $f$ as previously found.
The difference between planar and curved horizons is that the usual
horizon equation  $g_{tt} =0$ now becomes $f = -k\frac{L^2}{r^2}$. We
have not made a complete analysis of the structure of the new
spacetimes throughout $\mu$-$\lambda$ plane but let us make the
following preliminary remarks.

We can develop a qualitative picture of the solutions using the same
graphical approach as in the previous section. However, as well as
following the behaviour of the $\tilde{h}(f)$, we must now also keep
track of the critical value of $f$ where a horizon can form when
$k\ne0$, \ie $f_h\equiv-kL^2/r^2$. While the mass parameter $\omega^4$
controls how quickly $\tilde{h}(f)$ (and its roots) are shifting as $r$
varies, the rate of change in $f_h$ is controlled by the cosmological
constant scale $L$ while the direction is controlled by $k$. Our first
observation then is that if we have a black hole solution with $k=0$,
then for large masses $\omega\gg L$, we will always find the new
(curved) horizon equation with $r_h\sim\omega$. That is, in this
regime, the relevant root of $\tilde{h}(f)$ moves much more quickly as
$r$ decreases than $f_h$. Hence the root reaches $f=0$ at $r=\omega$
(as discussed for the planar horizons) while $|f_h|=L^2/\omega^2\ll1$
and so there should be a nearby solution for the horizon condition.
Therefore we expect that there are smooth black hole solutions with
spherical or hyperbolic horizons in all of the same regions of the
$\mu$-$\lambda$ plane where they were found for planar horizons, in the
previous section. However, in these regions, one may find that there is
a lower bound on the mass of these black hole solutions different from
$\omega^4=0$.

Let us consider small masses first for spherical horizons with $k=+1$
and $f_h=-L^2/r^2$. As $r$ decreases, the smallest positive root of
eq.~\reef{rewrite} moves to the left, approaching $f=0$ as
$r\rightarrow\omega$. At the same time, $f_h$ starts at zero at
$r=\infty$ and then moves to the left to negative values as $r$
decreases. Hence to form a horizon, the root must `catch up' to $f_h$.
If we tune $\omega^4$ to smaller and smaller values, slowing down the
rate at which the root moves, it becomes clear that it may never
coincide with $f_h$. For example, consider cases (d) with $p>0$, (e),
(f) and (h) in table \ref{table0}. In each of these cases, there will
be a value $f_0<0$ for which $\tilde{h}'(f_0)=0$, where the spacetime
develops a singularity as described in section \ref{planarbh} -- note
that this previous discussion does not change for $k\ne0$. Hence if
$f_h(\omega)=-L^2/\omega^2\le f_0$, then the root will not be able to
reach $f_h$ before hitting $f_0$. Hence in this situation, the
spacetime will contain a naked singularity. In cases (c), (d) with
$p<0$ and (g), $\tilde{h}(f)$ is monotonic for negative values of $f$.
However, now for small $r$ (\ie $r\ll\omega,L$) the root behaves as
$f\simeq -[\omega^4/(|\mu|\,r^{4})]^{1/3}$. Hence the root is growing
much more slowly than $f_h\propto r^{-2}$ in this regime and again it
becomes apparent that the root will never catch up to $f_h$. Hence the
solution will again have a naked singularity as $r\rightarrow0$. Thus
our final conclusion is that for spherical horizons with $k=+1$ there
will always be a lower (positive) bound on the mass parameter below
which no black hole solutions exist. This is, of course, qualitatively,
the same result as found for spherical black holes for Einstein gravity
with a negative cosmological constant \cite{hawkpage}. We have not
calculated the exact value of the lower bound but one must find
$\omega\sim L$, with the precise proportionality constant determined by
the gravitational couplings, $\lambda$ and $\mu$.

We might add that since $f_h<0$ with $k=+1$, the discussion given in
the previous section remains unchanged for the solutions asymptotic to
the ghosty vacua, the extra stable AdS vacuum in case (g) and the AdS
vacuum in case (c$_2$). That is, there will be no spherical black hole
solutions in any of these cases.

Now let us turn to considering small masses for hyperbolic horizons
with $k=-1$ and $f_h=L^2/r^2$. In this case, $f_h$ moves to the right
to positive values as as $r$ decreases and so more exotic possibilities
arise. For example, even if $\omega=0$ and the root does not move,
$f_h$ will increase and eventually coincide with the root. That is, a
horizon will appear even if the mass is set to zero and by continuity,
we must also find black holes with small negative masses. Note that
with $\omega^4$, the root moves to the right to values larger than the
initial $\fin$. Here, cases (e), (f) and (g), as well as the region
denoted (c$_1$), are distinguished because $\tilde{h}(f)$ is not
monotonic for positive values of $f>\fin$. Hence the negative mass
solutions will only contain a horizon if $f_h$ catches up to the root
before reaching the point where $\tilde{h}'=0$. This will set a
(negative) lower bound on the mass with $\omega^4\sim-L^4$ where again
the precise bound will be determined by $\lambda$ and $\mu$. Cases
(c$_0$), (d) and (h) with $k=-1$ are even more striking. For these
cases, $\tilde{h}(f)$ is monotonic for $f>\fin$ and so a singularity
only develops as $r\rightarrow0$ and $f\rightarrow\infty$. However, in
this regime, the root grows as $f\simeq
[|\omega^4|/(|\mu|\,r^{4})]^{1/3}$ while $f_h=L^2/r^2$. Hence the
solution will also contain a horizon with
$r_h^2\simeq|\mu|L^6/|\omega^4|$ for arbitrary negative values of
$\omega^4$. Hence we have found that in any of the parameter regimes
where planar black holes exist, there will be black holes with
hyperbolic horizons with negative masses. In cases (c$_1$), (e), (f)
and (g), there is a lower bound on how negative the mass can become but
in cases (c$_0$), (d) and (h), the hyperbolic black holes can have an
arbitrarily large negative mass. These results are not entirely
surprising given that hyperbolic black holes with negative masses exist
for for Einstein gravity with a negative cosmological constant
\cite{Roberto}, although there is a lower bound on the mass there.

We might briefly also consider the effect of setting $k=-1$ in the
cases where no planar black hole solutions could be found, \ie, the
solutions asymptotic to the ghosty vacua, the extra stable AdS vacuum
in case (g) and the AdS vacuum in case (c$_2$). It is clear that even
in these cases a smooth horizon forms with $\omega=0$ since
$f_h=L^2/r^2$ moves to positive values as $r$ decreases. Again by
continuity, black hole solutions will also exist for small positive and
negative values of $\omega^4$. As above, the largest root will grow as
$f\simeq [\omega^4/(\mu\,r^{4})]^{1/3}$ in a regime where
$r\rightarrow0$. Since the critical value $f_h$ grows more quickly, we
conclude that the solutions which asymptote to the ghosty vacua in
cases (e) and (f) (with $\mu>0$) will form a horizon for arbitrarily
large positive values of $\omega^4$. Similarly, the solutions which
asymptote to the second stable AdS vacuum in case (g) or to the AdS
vacuum in case (c$_2$) will form hyperbolic horizons for arbitrarily
large negative masses. Hence we find that exotic hyperbolic black holes
also exist in parameter regimes even where no planar black holes
formed.

\subsection{Higher dimensions} \label{higherD}

With the construction described in section \ref{Dcube}, we extended
quasi-topological gravity to higher dimensions $D\ge7$ -- recall that
our new curvature-cubed interaction did not affect the equations of
motion in $D=6$. The general action for $D\ge7$ (as well as $D=5$) is:
\beqa
 I &=& \frac{1}{16 \pi G_\mt{D}} \int \mathrm{d}^Dx \, \sqrt{-g}\, \left[
\frac{(D-1)(D-2)}{L^2} + R + \frac{\lambda L^2}{(D-3)(D-4)}\X_4 \right.
 \nonumber\\
&&\qquad\qquad\quad\left.-\frac{8(2D-3)
}{(D-6)(D-3)(3D^2-15D+16)}\mu L^4 \Z_\mt{D}\right] \labell{ActD}
 \eeqa
where $\X_4$ and $\Z_\mt{D}$ are given in eqs.~\reef{GBterm} and
\reef{result}, respectively. We choose the metric ansatz in
eq.~\reef{metric1}, which can describe $D$-dimensional black holes with
planar, spherical or hyperbolic horizons (for $k=0$, +1 and --1,
respectively). The coefficients of the action \reef{ActD} are chosen so
that substituting in this metric \reef{metric1} yields
 \beq
I=\frac{1}{16\pi G_\mt{D}} \int \mathrm{d}^Dx \, \frac{\left(D-2\right)
N\left(r\right)}{L^D}\, \left[ r^{D-1} \left( 1 - f +\lambda f^2 +\mu
f^3\right)\right]' \ . \labell{sippers}
 \eeq
The lapse must again be constant and a convenient choice is
$N(r)^2=1/\fin$ as described in section \ref{GBgrav} -- with the
implicit assumption that $\fin>0$. If we wish to consider the vacuum
solutions, $f(r)$ is also fixed to be constant with $f(r)=\fin$ where
these solutions are again determined by eq.~\reef{vacsol}. Hence the
various vacuum solutions are again distributed as described in table
\ref{table0}. Assuming $\fin>0$, the $D$-dimensional metric becomes
\beq ds^2 = -\left(k +\frac{r^2}{L^2}\fin\right) \frac{dt^2}{\fin}
+\frac{dr^2}{k+\frac{r^2}{L^2} \fin} + r^2 d\ell_k^2 \labell{vacmet1}
\eeq
where $d\ell^2_k$ is given in eq.~\reef{genan}. These solutions
correspond to a spherical ($k=+1$), flat ($k=0$) or hyperbolic ($k=-1$)
foliation of $AdS_\mt{D}$. From the $g_{rr}$ component, we see that the
radius of curvature of the $AdS_\mt{D}$ spacetime is
\beq
\tilde{L}^2=L^2/\fin\ .
\labell{vaccurv1}
\eeq

In considering the black hole solutions, the only difference from the
analysis for $D=5$ in the previous sections is that we replace:
${\omega^4}/{r^4} \rightarrow {\omega^{D-1}}/{r^{D-1}}$. In particular,
the latter substitution is made in $q$ in eq.~\reef{pnq}. Further,
eq.~\reef{rewrite} is replaced with
 \beq
\tilde{h}(f)\equiv\left(1-\frac{\omega^{D-1}}{r^{D-1}}\right)-f+\lambda
f^2 +\mu f^3 =0 \,.
 \labell{rewrite2}
 \eeq
The remainder of the analysis and the results in section \ref{planarbh}
carries over unaltered. In particular, table \ref{table0} correctly
describes the planar black hole solutions for $D\ge7$. Similarly, the
discussion of black holes with curved horizons in section \ref{curvhor}
remains largely unchanged. In certain cases, the discussion for small
masses referred to the behaviour of the root, now, of
eq.~\reef{rewrite2} as $r$ approaches zero. In the present case, this
behaviour changes to $f\simeq [\omega^{D-1}/(\mu\,r^{D-1})]^{1/3}$
while the behaviour of the critical value remains $f_h=L^2/r^2$. Hence
the root is grows more slowly than $f_h$ for $D=5$, at the same rate
for $D=7$ and more quickly for $D\ge8$. As a result, one finds, in
cases (c), (d) with $p<0$ and (g), that there is a positive lower bound
for the mass of spherical black holes for $D=5$ and 7 but the lower
bound is simply $\omega^{D-1}=0$ for $D\ge8$. Further, the conclusion
that hyperbolic black holes exits in cases (c$_0$), (d) and (h) with
arbitrarily large negative masses only applies for $D=5$. Similarly,
some of the details about the formation of hyperbolic horizons, in
regions of the coupling space where planar black holes do not exist,
change depending the spacetime dimension $D$.

As noted above, this discussion applies for $D\ge 7$, but in those
dimensions, one already has a cubic order theory in Lovelock gravity,
which also reproduces eq.~\reef{rewrite2}. Hence for $D\ge7$, the black
holes for quasi-topological gravity discussed here would be the same as
those in cubic Lovelock theory, as considered in \cite{jan3,Mann}.

\section{Black Hole Thermodynamics} \label{thermo}

We now turn to the thermodynamic properties of the black hole solutions
of quasi-topological gravity. Our focus will be on the planar ($k=0$)
black holes in five dimensions. The extension of these results to
curved horizons and higher dimensions is straightforward.

First, we use the standard approach to calculate temperature:
analytically continue the metric to Euclidean signature with $\tau =
-\imath t$ and periodically identify $\tau$ to produce an everywhere
smooth Euclidean section. Interpreting the period of $\tau$ as the
inverse temperature, we find
 \beq
T = \frac{1}{4\pi}\frac{r_h^2 f'|_{r_h}}{L^2 \sqrt{f_\infty}}\,,
\labell{temp0a}
 \eeq
which assumes the lapse is chosen as $N=1/\sqrt{\fin}$. For the planar
black holes, $r_h=\omega$ and further we will use $f(r=\omega)=0$. Then
we can evaluate the $f'|_{r_h}$ by differentiating the constraint
equation \reef{constr1} and evaluating the result at $r=\omega$. A
simple calculation yields $f'|_{\omega} = \frac{4}{\omega}$, giving:
 \beq
T = \frac{\omega}{\pi L^2 \sqrt{f_\infty}}\,. \labell{planarTh}
 \eeq

In passing we note that extending this calculation to general
dimensions and curved horizons yields
 \beqa
T&=&\frac{1}{4\pi\sqrt{f_\infty}}\,\left[\frac{r_h^2}{L^2}f'|_{r_h}
-\frac{2k}{r_h}\right]
\nonumber\\
&=& \frac{D-1}{4\pi\sqrt{f_\infty}}\,\left[\frac{\omega^{D-1}}{L^2
r_h^{D-2}}\,\frac{r_h^4}{r_h^4+2\lambda kL^2r_h^2-3\mu k^2L^4}
-\frac{2k}{D-1}\,\frac{1}{r_h}\right]\,, \labell{generalTh}
 \eeqa
where the precise location of the horizon $r_h$ must still be
determined for a spherical or hyperbolic horizon but, of course,
$r_h=\omega$ in the planar case ($k=0$).

Next, we calculate the entropy and energy densities of the black holes
following the Euclidean action approach, as already sketched for GB
theory in section \ref{GBgrav}. That is, we identify the Euclidean
action for the black hole solution, as the leading contribution to the
free energy, \ie $I_E\simeq F/T$.  Evaluating the Euclidean action
yields
 \beqa
I_{E}[T] &=& -\frac{1}{16 \pi G_\mt{5}} \int_{0}^{1/T} \mathrm{d}\tau
\int_{\omega}^{R_0}\mathrm{d}r \int \mathrm{d}^3 x \sqrt{g_E}
\left(\frac{12}{L^2}+R+\frac{\lambda L^2}{2}\chi_4
+\frac{7\mu}{4}\mathcal{Z}_5\right) \nonumber\\
&=& -\frac{V_3}{16\pi G_\mt{5}}\frac{1}{T
L^5\sqrt{f_\infty}}\left[r^4\left(3- 5f(r)+15\lambda f(r)^2-15\mu
f(r)^3\right) \right.
\labell{Eact2}\\
&&\qquad\qquad\left.+r^5\left(-1+6\lambda f(r)-9\mu f(r)^2 \right)f'(r)
\right]_\omega^{R_0}\,.\nonumber
 \eeqa
Here $V_3=\int \mathrm{d}^3x$ is the regulator volume for the (spatial)
gauge theory directions and as usual, we have set $N=1/\sqrt{\fin}$.
The final result is simplified using the constraint \reef{constr1} to
produce the following asymptotic expansion of $f(r)$:
 \beq
f\simeq f_\infty - \frac{\omega^4}{r^4}\frac{1}{\left(1 - 2\lambda
f_\infty -3\mu f_\infty^2\right)} +\cdots\,. \labell{asymf}
 \eeq
Keeping only the divergent and finite terms in the limit $R_0
\rightarrow \infty$, eq.~\reef{Eact2} then reduces to:
 \beq
I_{E}[T] = \frac{V_3}{8 \pi G_\mt{5}}\frac{\omega^4}{T
L^5\sqrt{f_\infty}}\left[\frac{R_0^4}{\omega^4}\fin\left(1
-6\lambda\fin +9\mu\fin^2 \right)
-\frac{1-4\lambda\fin+3\mu\fin^2}{1 - 2\lambda f_\infty -3\mu
f_\infty^2} \right]\,.\labell{eact3}
 \eeq
We remove the divergence in this expression by subtracting the action
for the AdS vacuum
 \beq
I_{E}^{0}[T'] = \frac{V_3}{8 \pi G_\mt{5}}\frac{R_0^4}{T'
L^5\sqrt{f_\infty}}\fin\left(1 -6\lambda\fin +9\mu\fin^2
\right)\,,\labell{eact4}
 \eeq
where $T'$ is chosen so that the periodicity of the AdS background
matches that of the black hole at the regulator surface $r=R_0$:
 \beq
\frac{1}{T'} = \frac{1}{T}\frac{\sqrt{f(R_0)}}{\sqrt{f_\infty}} \simeq
\frac{1}{T}\left(1-\frac{\omega^4}{2R_0^4 f_\infty\left(1 - 2\lambda
f_\infty -3\mu f_\infty^2\right)}\right)\,.\labell{tempp}
 \eeq
Combining these expressions yields a simple expression for the free
energy:
 \beq
F[T] = T\left(I_E[T] - I_E^0[T]\right) = -\frac{V_3 \omega^4}{16\pi
G_\mt{5} L^5\sqrt{f_\infty}} =-\frac{\left(\pi  \sqrt{f_\infty} L
\right)^3}{16\,G_\mt{5}}T^4\labell{freen}
 \eeq
where we have implicitly taken the limit $R_0\rightarrow\infty$ above.
As noted in section \ref{GBgrav}, for planar AdS black holes, we do not
have to account for the generalized Gibbons-Hawking term
\cite{genGibHawk} or the boundary counter-terms \cite{counter} in this
calculation of the free energy. Finally, to eliminate the regulator
volume, we work with the free energy density: ${\cal F}=F/V_3$. Then we
calculate the energy and entropy densities as
 \beqa
\rho &=& -T^2\frac{d\ }{dT}\left({\cal
F}/T\right)=\frac{3(\pi\sqrt{\fin} L)^3\, T^4}{16 G_\mt{5}}
\,,\labell{rhoands11}\\
s &=& -\frac{d{\cal F}}{dT}=\frac{\left(\pi\sqrt{\fin} L T\right)^3}{4
G_\mt{5}} = \frac{1}{4G_\mt{5}}\frac{\omega^3}{L^3}\,.\labell{sands11}
 \eeqa
Note that these expressions are `identical' to those appearing in
eq.~\reef{rhoands0} for GB gravity, however, $\fin$ implicitly depends
on the additional coupling $\mu$. We also comment that these results
obey the relation $\rho = \frac{3}{4}T s$, as expected for a
four-dimensional CFT at finite temperature.

\subsection{Noether Charge Approach to Entropy Density}

In this section, we verify that the entropy density in
eq.~\reef{sands11} matches the Wald entropy \cite{WaldEnt}. Of course,
we find agreement \cite{EntMatch} but the results for the Wald entropy
are also readily extended to higher dimensions and curved horizons.
Wald's prescription for the black hole entropy is
 \beq
S = -2 \pi \oint d^{D-2}x\sqrt{h}\ Y^{a b c d}\hat{\veps}_{a
b}\hat{\veps}_{c d} \qquad{\rm where}\ \ Y^{a b c d} =
\frac{\partial{\mathcal{L}}}{\partial R_{a b c d}} \labell{Waldformula}
 \eeq
$\mathcal{L}$ is the Lagrangian and $\hat{\veps}_{a b}$ is the binormal
to the horizon. For the static black holes considered here, $Y=Y^{a b c
d}\hat{\veps}_{a b}\hat{\veps}_{c d}$ is constant on the horizon and so
the entropy is given simply as
 \beq
S=-2\pi Y A,
 \eeq
where $A = \oint d^{D-2}x\sqrt{h}$ is the `area' of the horizon. Let us
divide up the terms in the action \reef{ActD} according to their powers
of the curvature tensor: the Einstein term, the GB interaction and the
quasi-topological interaction. (Of course, one also has the
cosmological constant term but it does not contribute in Wald's formula
\reef{Waldformula}.) Following the above prescription, as usual, the
Einstein term yields $Y_1=-1/(8\pi G_\mt{D})$ and the resulting entropy
is the expected Bekenstein-Hawking entropy $S=A/(4G_\mt{D})$. Applying
this formalism to GB terms, we find
 \beq
Y_2 =-\frac{1}{4\pi G_\mt{D}}\frac{\lambda\,L^2}{(D-3)(D-4)} \left(R -
2\left(R^{t}{}_{t}+R^{r}{}_{r}\right)+2R^{t r}{}_{t r}\right)\,,
\labell{GBWaldent}
 \eeq
where this expression applies for a general static black hole metric.
One can think that the tensor components above are presented in an
orthonormal frame or alternatively in a coordinate frame, as long as
the indices are in precisely the raised and lowered positions as shown.
Integrating this over the horizon gives the following contribution to
the entropy as
 \beqa
S_2 &=& \frac{A}{2G_\mt{D}}\frac{\lambda\,L^2}{(D-3)(D-4)}\left(R -
2\left(R^{t}_{\,\,t}+R^{r}_{\,\,r}\right)+2R^{t r}_{\,\,\,\,t
r}\right)\nonumber\\
&=&\frac{A}{4G_\mt{D}}\frac{D-2}{D-4}\left( -2\lambda\,f(r_h)\right)\,.
 \labell{GBWaldent2}
 \eeqa
where we are using the general metric \reef{metric1} with $N^2=1/\fin$
to evaluate the second line. Finally turning to the quasi-topological
contribution in the action, we find
 \beqa
Y_{3} &=& \frac{1}{4\pi G_\mt{D}}\frac{8(2D-3)\,\mu\,L^4
}{(D-6)(D-3)(3D^2-15D+16)}\left[\frac{3\c1}{2}\left(R^{tm}{}_{t n}R^{r
n}{}_{rm}-R^{tm}{}_{r n}R^{r}{}_m{}_t{}^n\right)\right.
\labell{qtentropy}\\
&&\qquad\quad+3\c2R^{t r m n}R_{t r m n}+\c3\left(R^{t r}{}_{t
m}R_r{}^{m}-R^{t r}{}_{r m} R_t{}^{m}+\frac{1}{4}\left(R_{m n p r}R^{m
n p r}+R_{m n p t}R^{m n pt}\right)\right)
 \nonumber\\
&&\qquad\quad+\c4\left(2R\,R^{t r}{}_{t r}+\frac{1}{2}R_{m n p q}R^{m n
p q}\right)+\frac{c_5}{2}\left(R^{t}{}_{t}R^{r}{}_{r}
-R^t{}_{r}R^r{}_{t}+R^{r}{}_{m r n}R^{m n}+R^t{}_{m t n}R^{m n}\right)
\nonumber\\
&&\qquad\quad\left.+\frac{3}{4}\c6 \left(R^{r m}R_{r m} +R^{t m}R_{t
m}\right)+\frac{\c7}{2}\left(R_{mn}R^{mn}+R\left(R^r{}_{r}+
R^{t}{}_{t}\right)\right)+\frac{3}{2}\c8 R^2 \right]\,.\nonumber
 \eeqa
We have left the coefficients arbitrary above but it is understood that
they are to be fixed as in eq.~\reef{table2}. Now integrating over the
horizon yields
 \beqa
S_3 &=& -\frac{A}{2G_\mt{D}}\frac{8(2D-3)\,\mu\,L^4
}{(D-6)(D-3)(3D^2-15D+16)}\left[\frac{3\c1}{2}\left(R^{tm}{}_{t n}R^{r
n}{}_{rm}-R^{tm}{}_{r n}R^{r}{}_m{}_t{}^n\right)\right.
\nonumber\\
&&\qquad\quad+3\c2R^{t r m n}R_{t r m n}+\c3\left(R^{t r}{}_{t
m}R_r{}^{m}-R^{t r}{}_{r m} R_t{}^{m}+\frac{1}{4}\left(R_{m n p r}R^{m
n p r}+R_{m n p t}R^{m n pt}\right)\right)
 \nonumber\\
&&\qquad\quad+\c4\left(2R\,R^{t r}{}_{t r}+\frac{1}{2}R_{m n p q}R^{m n
p q}\right)+\frac{c_5}{2}\left(R^{t}{}_{t}R^{r}{}_{r}
-R^t{}_{r}R^r{}_{t}+R^{r}{}_{m r n}R^{m n}+R^t{}_{m t n}R^{m n}\right)
\nonumber\\
&&\qquad\quad\left.+\frac{3}{4}\c6 \left(R^{r m}R_{r m} +R^{t m}R_{t
m}\right)+\frac{\c7}{2}\left(R_{mn}R^{mn}+R\left(R^r{}_{r}+
R^{t}{}_{t}\right)\right)+\frac{3}{2}\c8 R^2 \right]\nonumber\\
&=&\frac{A}{4G_\mt{D}}\frac{D-2}{D-6}\left( -3\mu\,f(r_h)^2\right)\,.
\labell{GenR3Entropy}
 \eeqa
Combining all of these expressions, we arrive at the Wald entropy for
quasi-topological gravity:
 \beq
S= \frac{A}{4G_\mt{D}}\left(1-\frac{2(D-2)}{D-4}\lambda\,f(r_h)
-\frac{3(D-2)}{D-6}\mu\,f(r_h)^2 \right)\,.\labell{R3S}
 \eeq
Evaluating this expression on a planar horizon yields the simple
result, $S=A/(4G_\mt{D})$, \ie, the higher curvature contributions
vanish on planar horizons. If we divide by the regulator volume, this
yields the entropy density:
 \beq
s=\frac{S\ }{V_{D-2}}=\frac{\omega^{D-2}}{4G_\mt{D}\,L^{D-2}}\,,
 \labell{entdens55}
 \eeq
which, of course, agrees with the result in eq.~\reef{sands11} for
$D=5$. For the case of black holes with curved horizons, as in section
\ref{higherD}, we find that the entropy is given by
 \beq
S_k = \frac{A}{4G_\mt{D}}\left(1+\frac{2(D-2)}{D-4}\lambda\,k
\frac{L^2}{r_h^2} -\frac{3(D-2)}{D-6}\mu\,k^2 \frac{L^4}{r_h^4}
\right)\,.\labell{R3Scruv}
 \eeq

\section{Equations of Motion} \label{eom}

We have found that the equations of motion for quasi-topological
gravity take an incredibly simple form with the ansatz \reef{metric1}
for a static AdS black hole -- \eg see eq.~\reef{sippers}. In this
section, we would like to investigate the equations of motion in
greater generality to gain some further insight into this simplicity.
In particular, we will see the linearized equations of motion of
graviton fluctuations also exhibit a certain simplicity. We have
already argued that the new cubic interactions constructed in section
\ref{new} do not have a topological origin -- see appendix
\ref{app-topo}. Hence this cannot be the source of the simplicity noted
above.

Let us begin with the cubic-curvature interactions in eq.~\reef{ZZ}.
First we set $\c{11}=0=\c{13}$ and then find the general contribution
these terms would make to the metric equations of motion:
\beqa \frac{1}{\sqrt{-g}}\frac{\delta I}{\delta g^{a b}} &=& \c8
\left(-3R^2 R_{a b}  +3\,(R^2)_{;ab}\right) +\c7\left(-R_c{}^d R_d{}^c
R_{a b} -2 R R_{a c}R^c{}_b +(R_c{}^d R_d{}^c)_{;a b}-\Box\left(R R_{a
b}\right)\right.
 \nonumber\\
&&\quad \left.+2(R R^c{}_{\left(a \right.})_{\left.;b\right)c}
\right)+\c6\left(-3R_{a c}R^c{}_d R^d{}_b -\frac{3}{2}\Box(R_{a
c}R^c{}_b) +3(R^c{}_d R^d{}_{\left(b\right.})_{\left. ;a \right) c}
\right)
\nonumber\\
&&\quad+\c5\left(-3R_{c \left(a\right.}R_{\left.b\right) d}{}^c{}_eR
R^{d e}+2(R_{\left(a\right.| c}{}^d{}_e R^{c
e})_{|\left.;b\right)d}-\Box(R_{a c  b d}R^{c
d})+(R_{\left(a\right.}{}^c R_{\left.b\right)}{}^d)_{;c d}\right.
\nonumber\\
&&\quad \left.-(R^{c d}R_{a b})_{;c d} \right)+ \c4\left(-2R R_{a c d
e}R_{b}{}^{c d e}-R_{a b}R_{c d e f}R^{c d e f}+(R_{c d e f}R^{c d e
f})_{;a b}\right.
\nonumber\\
&&\quad\left.+4(R R_a{}^{c d}{}_{b})_{;c d} \right)+\c3\left(-2R_{a c
d}{}^e R_{b}{}^{c d f} R_{e f}-R^{d e f c}R_{d e
f\left(a\right.}R_{\left.b\right) c}+(R_{c d e}{}^f R^{c d
e}{}_{\left(a\right.})_{\left.;b\right) f}\right.
\nonumber\\
&&\quad\left.+\frac{1}{2}\Box(R^{c d e}{}_{a}R_{c d e b})+2(R^{e d
c}{}_{\left(a\right.}R_{\left.b\right)e})_{;c d}+2(R^e{}_{(a b)}{}^c
R_e{}^d)_{;c d} \right)+ \c2\left(R_{a c}{}^{d e}R_{d e}{}^{f g}R_{f g
b}{}^c\right.
\nonumber\\
&&\quad\left.+6(R_{\left(a\right.}{}^{c e f}R^d{}_{\left.b\right) e
f})_{;c d} \right)+\c1\left(-3R_{c d e f}R^{c g e}{}_aR_{f g b}{}^d
+3(R_e{}^d{}_{f \left(b\right.}R_{\left.a\right)}{}^{e c f})_{;c
d}\right.
\nonumber\\
&&\quad\left.-3(R_{\left(a\right.}{}^e{}_{\left.b\right)}{}^f
R_e{}^d{}_f{}^c)_{;d c} \right)+g_{a
b}\left(\c8\left(\frac{1}{2}R^3-3\,\Box(R^2)\right)+\c7\left(\frac{1}{2}R
R_c{}^d R_d{}^c -(R R^{c d})_{;c d}\right.\right.
\nonumber\\
&&\quad\left.-\Box(R_c{}^d R_d{}^c) \right)+\c6\left(\frac{1}{2}R_c{}^d
R_d{}^e R_e{}^c-\frac{3}{2}(R^c{}_e R^{e d})_{;c d}
\right)+\c5\left(\frac{1}{2}R^{c d}R_{c e d f}R^{e f}\right.
\nonumber\\
&&\quad\left. -(R_e{}^c{}_f{}^d R^{e f})_{;c d}
\right)+\c4\left(\frac{1}{2}R R_{c d e f}R^{c d e f} -\Box(R_{c d e
f}R^{c d e f}) \right)+\c3\left(\frac{1}{2}R^{c d e f}R_{c d e}{}^g
R_{f g} \right.
\nonumber\\
&&\quad\left.\left.-\frac{1}{2}(R^{d e g c}R_{d e g}{}^f)_{;e f}
\right) +\frac{1}{2}\c2 R_{c d}{}^{e f}R_{e f}{}^{g h}R_{g h}{}^{c
d}+\frac{1}{2}\c1 R_c{}^d{}_e{}^f R_d{}^g{}_f{}^h
R_g{}^c{}_h{}^e\right)\ . \labell{R3var1}
 \eeqa
Since we have eliminated the terms involving derivatives of the
curvature from the action by setting $\c{11}=0=\c{13}$, the above
contributions contain at most terms with four derivatives of the
metric, such as in  $\Box\left(R R_{a b}\right)$. This may be
misleading, however, since we have made no attempt to simplify this
expression using, \eg the Bianchi identities. A useful check is to
choose the coefficients above so that eq.~\reef{ZZ} corresponds to the
six-dimensional Euler density \reef{euler6}, \ie choose $\c1=-2\c2$,
$\c2=4$ and the remaining coefficients as in eq.~\reef{table1}. In this
case, we were able to verify that any terms involving derivatives of
curvatures can be eliminated and the expected field equations involving
only factors of the curvature were reproduced -- \eg see
\cite{Mann,Cubevar}. Indeed Lovelock's general discussion
\cite{lovelock} dictates that $\X_6$ is the only gravity Lagrangian
which is cubic in curvatures for which the equations of motion do not
contain any derivatives of curvatures. Hence one must expect equations
of motion that include terms with derivatives of curvatures for any
other choice of the coefficients $\c{i}$ and, in particular, for
quasi-topological gravity with $\c{i}$ as in eq.~\reef{table2}. While
we still made some effort to `tidy up' the derivative terms in
eq.~\reef{R3var1}, we did not produce any particularly illuminating
results.

As a next step, we examine the linearized equations of motion for a
graviton perturbation in quasi-topological gravity. Hence we fixed the
coefficients as in eq.~\reef{table2} and then substitute into the above
expression \reef{R3var1}: $g_{ a  b}= g_{a b}^\mt{[0]}+h_{ a b}$ where
$g_{a b}^\mt{[0]}$ is a solution of the full equations of motion.
Again, the resulting expression of a generic fluctuation is rather
complicated and so to proceed further, we restrict our attention to
transverse traceless gauge with $\nabla^a h_{a b} =0$) and $h^a{}_a=0$.
This choice simplifies the result somewhat and the four-derivative
contribution is proportional to:
 \beqa
&&(D-4)R^{ cd ef}\,h_{ de; cf\left( a b\right)}+R^{ cd } \left(\Box h_{
cd }\right){}_{\!;\left( a  b\right)}-2R^{ cd }\left(\Box h_{
c\left(\right. a }\right){}_{\!; \left.b\right) d}+\frac{2}{D-2}R^{ cd
}\left(\Box h_{ a b}\right){}_{\!; cd }
\qquad\quad\nonumber\\
&&\quad+2\left(\Box h^{ c}{}_{\left(\right. a }\right){}^{\!; de}R_{|
cd e| b\left.\right)} +g_{ a  b}\left(\Box h_{ cd }\right)_{;e f}R^{ ce
d f}+2R_{ c\left( a \right.}\,\Box^2h^{ c}{}_{\left. b \right)}
  -\frac{1}{2}g_{ a  b}\,R_{ cd }\,\Box^2h^{ cd }\nonumber\\
&&\quad+\left(D-3\right)\Box^2h^{ cd }R_{ c a  d
b}-\frac{R}{\left(D-2\right)}\Box^2h_{ a b}\,,\labell{fourder}
  \eeqa
where we use the standard notation $T_{(ab)}=\frac{1}{2}\left( T_{ab} +
T_{ba}\right)$. Hence despite the reduction in the number of terms, we
see that for general backgrounds, the linearized equations of motion
for (physical) gravitons include four-derivative contributions.

Of course, if we were considering gravitons propagating in flat space,
these contributions would all vanish because $R_{abcd}=0$ in the
background spacetime. The same result would apply for any interactions
which are cubic in curvatures for a flat background. However, in the
present context, it is natural to consider gravitons propagating in an
AdS$_D$ background. In this case, of course, the background curvature
is nonvanishing and so one might expect these terms \reef{fourder} will
still appear in the linearized equations of motion. However, further
simplifications can be expected since AdS$_D$ is a maximally symmetric
spacetime with\footnote{Note that here we are distinguishing
$\tilde{L}$, the curvature scale of the AdS background, from $L$, the
AdS length scale appearing in the action. In particular, recall that in
the vacuum solutions above, we found $\tilde{L}=L/\sqrt{f_\infty}$.}
 \beq
R_{ a b c d}=-\frac{1}{\tilde{L}^2}\,\left(g_{a c}\,g_{b d}-g_{a d}\,
g_{b c}\right)\,. \labell{MSymRie}
 \eeq
Remarkably, one finds that upon substituting this background curvature
into eq.~\reef{fourder}, all of the remaining four-derivative terms
cancel! Further, because the background curvature \reef{MSymRie} is
covariantly constant, there are no nontrivial terms with only three
derivatives acting on the graviton. Therefore, in quasi-topological
gravity, the linearized graviton equation in an AdS$_D$ background is
only a second-order equation.

With this result in hand, we next construct the linearized equation of
motion for the graviton in the AdS$_D$ vacuum solutions. Hence we must
consider the full action \reef{ActD} for quasi-topological gravity,
including the cosmological constant, Einstein and Gauss-Bonnet terms as
well. The equation of motion for the transverse traceless graviton can
then be written as:
%\footnote{This result was deduced by examining
%various polarizations in the $D$-dimensional extension of the vacuum
%solution \reef{vacmet} for values of $D$.}
 \beq
-\frac{1}{2}\left(1-2\lambda f_\infty-3\mu f_\infty^2
\right)\left[\nabla^2 h_{a b} +\frac{2\fin}{L^2}\, h_{a b}\right]=8\pi
G_\mt{D}\,\hat{T}_{ab}\,. \labell{Boxh}
 \eeq
We have added a stress-tensor on the right-hand side, as might arise
from minimally coupling the metric to additional matter fields or from
quadratic or higher order contributions in the graviton. The second
bracketed factor on the right-hand side is the standard Einstein
equation of motion for gravitons in an AdS background with curvature
$\tilde{L}=L/\sqrt{\fin}$ \cite{standard}. One can recognize the first
bracketed factor as the slope of the cubic equation \reef{vacsol}
determining $\fin$, \ie see eq.~\reef{vacslope}. Hence this slope
determines the sign of the graviton propagator or alternatively the
sign of the coupling of the graviton to the stress tensor. We see that
the appropriate sign for a well-behaved graviton is negative since this
factor reduces to --1 when $\lambda=0=\mu$. Hence as discussed in
section \ref{planarbh}, the AdS vacua are only stable when the slope is
negative while the graviton is a ghost in AdS backgrounds where the
slope is positive.

While restricting to transverse traceless gauge is a convenient
simplification to determine the linearized equations in an AdS
background, we can do better. A straightforward argument shows that the
above gauge-fixed equations of motion \reef{Boxh} extend to the full
linearized Einstein equations \cite{standard}:\footnote{We would like
to thank Miguel Paulos for discussions on this point and confirming
that several trial perturbations satisfied eq.~\reef{fullEin} with
Mathematica.}
 \beqa
&&-\frac{1}{2}\left(1-2\lambda f_\infty-3\mu f_\infty^2
\right)\left[\nabla^2h_{a b} + \nabla_a\nabla_b\, h_c{}^c-
\nabla_a\nabla^ch_{cb}- \nabla_b\nabla^ch_{ca}
 \vphantom{\frac{2\fin}{L^2}} \right.
 \labell{fullEin}\\
&&\qquad\qquad\left.-g^\mt{[0]}_{ab}\left(\nabla^2
h_c{}^c-\nabla^c\,\nabla^dh_{cd}\right) +\frac{2\fin}{L^2}\, h_{a
b}-\frac{(D-3)\fin}{L^2}\,g^\mt{[0]}_{ab}\, h_c{}^c\right]=8\pi
G_\mt{D}\,\hat{T}_{ab}\,, \nonumber
 \eeqa
where $g^\mt{[0]}_{ab}$ is the background AdS metric. We know that the
full linearized equations come from a covariant expression and hence
they must invariant under the `gauge' transformations: $\delta h_{ab}=
\nabla_{\!a}\,\varepsilon_b +\nabla_{\!b}\,\varepsilon_a$. Now the
linearized Einstein equations \reef{fullEin} are certainly invariant
under these transformations and reduce to eq.~\reef{Boxh} upon fixing
to transverse traceless gauge. However, there may be additional
contributions which are both gauge invariant and completely vanish for
transverse traceless modes. For example, the full equations may include
an additional contribution proportional to
 \beq
g^\mt{[0]}_{ab}\left(\nabla^2 h_c{}^c-\nabla^c\,\nabla^dh_{cd}
+\frac{(D-1)\fin}{L^2}\, h_c{}^c\right)\,,
 \labell{adder}
 \eeq
which is gauge invariant but would not contribute in eq.~\reef{Boxh}.
In this particular case, it is easy to argue that such a contribution
could not arise from the variation of an action (quadratic in
$h_{ab}$). However, given the action \reef{ActD} for quasi-topological
gravity, another approach is to evaluate the quadratic action (using
Mathematica) and subsequently examining the equations of motion for
some specific trial perturbations which are not transverse or
traceless. In every case considered, the latter equations match
precisely the results expected from eq.~\reef{fullEin}. This clearly
shows that there are no additional terms of the form given in
eq.~\reef{adder} but more importantly that there are no additional
four-derivative contributions in the full linearized equations without
gauge-fixing. The fact that gravitons propagating in an AdS background
simply obey the same equations of motion as in Einstein gravity plays
an important role in understanding the holographic properties of
quasi-topological gravity \cite{qthydro}.

\section{Discussion} \label{discuss}

In section \ref{new}, we have constructed a new gravitational action
which includes terms cubic in the curvature. Our construction was
motivated by the simple equations \reef{quadd} arising to determine the
black hole solutions in GB gravity. We were able to reproduce a similar
structure \reef{constr1} for our new theory. We wish to emphasize how
remarkable this result is. In section \ref{eom}, we showed that the
full equations are fourth order in derivatives. However, in section
\ref{BHs}, we found that once the geometry of the horizon is fixed, the
static black hole solutions are fixed by a single integration constant!
It seems that the symmetry imposed on the background geometry must play
an important role in producing the simplicity of these solutions. While
our approach was to substitute the ansatz into the action, it would be
interesting to work directly with the equations of motion and formalize
this result in terms of a `Birkhoff theorem.' It would also be
interesting to see to what extent this simplicity extends to spinning
or electrically charged black holes in quasi-topological gravity.

We also saw that the linearized equations of motion for gravitons
propagating in the AdS backgrounds reduced to the same second order
equations as for Einstein gravity in section \ref{eom}. There we saw
more or less directly that this simplification comes about due to the
maximal symmetry of the AdS spacetime. On the other hand, we should not
expect that the four-derivative contributions \reef{fourder} to the
equations of motion cancel in the black hole backgrounds. Hence the
quasinormal spectrum of the black hole solutions should be studied in
detail. In particular, this spectrum may reveal that these solutions
are unstable for certain values of the gravitational couplings.

The simplicity of the graviton equations in an AdS background has the
interesting consequence that the standard holographic rules apply in
matching the metric fluctuations to the stress tensor of the CFT. In a
general higher derivative theory, implicitly the graviton would be
matched with some higher dimension operator, as well as the stress
tensor \cite{hofman}. This interpretation arises because the higher
derivative equations allow the metric fluctuations to have more that
the standard asymptotic behaviour in the AdS geometry. In any event,
this complication is evaded in quasi-topological gravity and so the
effect of the higher derivative gravitational terms will only be felt
in the higher $n$-point couplings of the stress tensor.

Having motivated the construction of quasi-topological gravity by
considerations of the AdS/CFT correspondence, one might ask how the
universality class of dual CFT's has been expanded. The gravitational
theory is defined by three independent dimensionless parameters:
$\lambda$, $\mu$ and $L^{D-2}/G_\mt{D}$. In general, the three-point
function of the stress tensor of a CFT in four or higher dimensions is
also characterized by three independent (dimensionless) parameters
\cite{ozzy}. This match in the counting  of these parameters is not a
coincidence, as the discussion of \cite{HM} that holographically
modelling the full range of these CFT parameters requires the
introduction of curvature-squared and curvature-cubed interactions in
the bulk gravity theory. We make precise the mapping between the
gravitational couplings and the dual CFT parameters, as well as
exploring other holographic aspects of quasi-topological gravity in
\cite{qthydro,ctheorem}.

Another natural extension of this work is to consider analogous
gravitational interactions with higher powers of the curvature. Of
course, Lovelock gravity provides an infinite sequence of
(curvature)$^n$ interactions which still allow for a certain
calculational control in the gravitational theory. However, because of
their topological origin, these Lovelock terms only contribute to the
equations of motion for $D\ge 2n+1$, \ie in the context of the AdS/CFT
correspondence, for dual CFT's with $d\ge2n$. However, our expectation
is that our construction of five-dimensional quasi-topological gravity
with curvature-cubed interactions can be extended general
(curvature)$^n$ interactions and this has been verified by some
preliminary calculations \cite{moreunpub}. In fact, our conjecture is
that in $D\ge 2n+1$, an independent interaction can be constructed for
each independent scalar contraction of (Weyl tensor)$^n$. For $D<
2n+1$, Schouten identities will reduce the number of independent
interactions, as seen in the present analysis of the curvature-cubed
interactions. However, our understanding of this issue remains
incomplete, as we are still uncertain as to why there was no effective
curvature-cubed interaction in six dimensions. In any event, better
understanding the number of independent quasi-topological gravity terms
with higher powers of the curvature will be an interesting direction of
study. Important new insights into this question were given in
refs.~\cite{newer,newer2}. There it was shown that the curvature-cubed
interactions constructed here can be simply expressed in terms of
scalar contractions of the Weyl tensor combined with the
six-dimensional Euler density. Their construction immediately
generalizes to an infinite family of higher curvature interactions with
similar properties.

\acknowledgments We thank Ted Jacobson, Barak Kol, Robb Mann, Miguel
Paulos and Aninda Sinha for helpful discussions and useful comments. We
also thank Jorge Escobedo for his help in preparing the figures and for
proofreading the paper. RCM would also like to thank the KITP and the
Weizmann Institute for hospitality at various stages of this project.
Research at the KITP is supported by the National Science Foundation
under Grant No. PHY05-51164. Research at Perimeter Institute is
supported by the Government of Canada through Industry Canada and by
the Province of Ontario through the Ministry of Research \& Innovation.
We also acknowledge support from an NSERC Discovery grant and funding
from the Canadian Institute for Advanced Research.

\appendix

\section{A New Topological Invariant?}\label{app-topo}

In section \ref{Dcube}, our construction produced an nontrivial
interaction \reef{result} for any number of dimensions $D\ge5$.
However, we noted $\Z_6$, the six-dimensional expression, did not
contribute to the equations of motion for the black hole metric
\reef{metric0} (extended to $D=6$). In particular, this also means that
the the AdS vacua are unaffected by the addition of $\Z_6$ to the
gravitational action. This behaviour is reminiscent of Lovelock gravity
where, for example, the Euler density $\X_6$ provides a nontrivial
gravitational interaction for $D\ge7$ but because of its topological
origin, it leaves the equations of motion unaffected for $D\le6$. Hence
one might be tempted to believe that $\Z_6$ yields another topological
invariant in six dimensions. However, in the following, we demonstrate
that $\int d^6x\sqrt{g}\,\Z_6$ is not a topological invariant by
explicitly evaluating this expression for certain specific
six-dimensional geometries.

As our first test, we evaluate this expression on a deformed six-sphere
with metric:
 \beq
ds^2 = R^2\,\left[\,d\theta^2 +\sin^2\theta\,\left(1+a \,
\sin^2\theta\right)^n\,d\Omega_5^2\,\right] \labell{warp}
 \eeq
where $n$ (implicitly an integer) and $a$ are constants defining the
deformation away from the round six-sphere. We then find:
 \beq
\int_{_{{S}^6}}\sqrt{g}\,\,\Z_6 = \frac{544}{3}\pi^3\,, \qquad
\int_{_{{S}^6}}\sqrt{g}\,\,\X_6 = 768 \pi^3\,. \labell{int99a}
 \eeq
where we have normalized $\X_6$ as in eq.~\reef{euler6}. Hence we see
that both of these results are independent of the deformation
parameters. Of course, for $\X_6$, this occurs because the integrated
expression is a topological invariant. While again suggestive for
$\Z_6$, this result is by no means conclusive and hence we consider a
further test.

Next we consider the following metric in which the spheres in the
direct product $S^2\times S^4$ are deformed:
 \beqa
ds^2 &=&R^2\,\left[ d\theta^2 +\sin^2\theta\,\left(1+a \,
\sin^2\theta\right)^2\,d\phi^2\right]
 \labell{warp2}\\
&&\qquad + L^2\,\left[d\tilde\theta^2 +\sin^2\tilde\theta\,\left(1+b \,
\sin^2\tilde\theta\right)^2\,d\Omega_3^2\right]
 \nonumber
 \eeqa
where the deformation is characterized by the constants $a$ and $b$. In
this case, we find:
 \beq
\int_{_{S^2\times S^4}}\sqrt{g}\,\,\Z_6 = F(a,b,R/L)\,, \qquad
\int_{_{S^2\times S^4}}\sqrt{g}\,\,\X_6 = 1536 \pi^3\,. \labell{int99b}
 \eeq
where $F(a,b,R/L)$ is a complicated (and not particularly illuminating)
function of both deformation parameters and the relative radius of
curvature of the two spheres. Hence this result makes clear that $\Z_6$
does not yield a topological invariant.


\begin{thebibliography}{99}

\bibitem %[kss]
 {kss} P.~Kovtun, D.~T.~Son and A.~O.~Starinets,
  ``Viscosity in strongly interacting quantum field theories from black hole
  physics,''
  Phys.\ Rev.\ Lett.\  {\bf 94}, 111601 (2005)
  [arXiv:hep-th/0405231];\\
  %%CITATION = PRLTA,94,111601;%%
P.~Kovtun, D.~T.~Son and A.~O.~Starinets,
  ``Holography and hydrodynamics: Diffusion on stretched horizons,''
  JHEP {\bf 0310}, 064 (2003)
  [arXiv:hep-th/0309213].
  %%CITATION = JHEPA,0310,064;%%

\bibitem %[einstein]
 {einstein} A.~Buchel and J.~T.~Liu,
  ``Universality of the shear viscosity in supergravity,''
  Phys.\ Rev.\ Lett.\  {\bf 93}, 090602 (2004)
  [arXiv:hep-th/0311175];\\
  %%CITATION = PRLTA,93,090602;%%
A.~Buchel,
  ``On universality of stress-energy tensor correlation functions in
  supergravity,''
  Phys.\ Lett.\  B {\bf 609}, 392 (2005)
  [arXiv:hep-th/0408095];\\
  %%CITATION = PHLTA,B609,392;%%
P.~Benincasa, A.~Buchel and R.~Naryshkin,
  ``The shear viscosity of gauge theory plasma with chemical potentials,''
  Phys.\ Lett.\  B {\bf 645}, 309 (2007)
  [arXiv:hep-th/0610145];\\
  %%CITATION = PHLTA,B645,309;%%
D.~Mateos, R.~C.~Myers and R.~M.~Thomson,
  ``Holographic viscosity of fundamental matter,''
  Phys.\ Rev.\ Lett.\  {\bf 98}, 101601 (2007)
  [arXiv:hep-th/0610184];\\
  %%CITATION = PRLTA,98,101601;%%
K.~Landsteiner and J.~Mas,
  ``The shear viscosity of the non-commutative plasma,''
  JHEP {\bf 0707}, 088 (2007)
  [arXiv:0706.0411 [hep-th]];\\
  %%CITATION = JHEPA,0707,088;%%
N.~Iqbal and H.~Liu,
  ``Universality of the hydrodynamic limit in AdS/CFT and the membrane
  paradigm,''
  Phys.\ Rev.\  D {\bf 79}, 025023 (2009)
  [arXiv:0809.3808 [hep-th]].
  %%CITATION = PHRVA,D79,025023;%%

\bibitem %[string4]
 {string4} For example, see:\\
A.~Buchel, J.~T.~Liu and A.~O.~Starinets,
  ``Coupling constant dependence of the shear viscosity in N=4 supersymmetric
  Yang-Mills theory,''
  Nucl.\ Phys.\  B {\bf 707}, 56 (2005)
  [arXiv:hep-th/0406264];\\
  %%CITATION = NUPHA,B707,56;%%
P.~Benincasa and A.~Buchel,
  ``Transport properties of N = 4 supersymmetric Yang-Mills theory at  finite
  coupling,''
  JHEP {\bf 0601}, 103 (2006)
  [arXiv:hep-th/0510041];\\
  %%CITATION = JHEPA,0601,103;%%
A.~Buchel,
  ``Shear viscosity of boost invariant plasma at finite coupling,''
  Nucl.\ Phys.\  B {\bf 802}, 281 (2008)
  [arXiv:0801.4421 [hep-th]];
  %%CITATION = NUPHA,B802,281;%%
 ``Resolving disagreement for $\eta/s$ in a CFT plasma at finite coupling,''
  Nucl.\ Phys.\  B {\bf 803}, 166 (2008)
  [arXiv:0805.2683 [hep-th]];\\
  %%CITATION = NUPHA,B803,166;%%
R.~C.~Myers, M.~F.~Paulos and A.~Sinha,
  ``Quantum corrections to $\eta/s$,''
  Phys.\ Rev.\  D {\bf 79}, 041901 (2009)
  [arXiv:0806.2156 [hep-th]].
  %%CITATION = PHRVA,D79,041901;%%


\bibitem %[string]
 {string} Y.~Kats and P.~Petrov, ``Effect of curvature
    squared corrections in AdS on the viscosity of the dual gauge
    theory,''
  JHEP {\bf 0901}, 044 (2009)
  [arXiv:0712.0743 [hep-th]];\\
  %%CITATION = JHEPA,0901,044;%%
A.~Buchel, R.~C.~Myers and A.~Sinha,
  ``Beyond $\eta/s = 1/4\pi$,''
  JHEP {\bf 0903}, 084 (2009)
  [arXiv:0812.2521 [hep-th]];\\
  %%CITATION = JHEPA,0903,084;%%
R.~C.~Myers, M.~F.~Paulos and A.~Sinha,
 ``Holographic Hydrodynamics with a Chemical Potential,''
  JHEP {\bf 0906}, 006 (2009)
  [arXiv:0903.2834 [hep-th]].
  %%CITATION = JHEPA,0906,006;%%

\bibitem %[EtasGB]
 {EtasGB} M.~Brigante, H.~Liu, R.~C.~Myers, S.~Shenker and
    S.~Yaida, ``Viscosity Bound Violation in Higher Derivative Gravity,''
  Phys.\ Rev.\ D {\bf 77} (2008) 126006
  [arXiv:htp-th/0712.0805];
  %%CITATION = PHRVA,D77,126006;%%
``The Viscosity Bound and Causality Violation,''
  Phys.\ Rev.\ Lett.\  {\bf 100}, 191601 (2008)
  [arXiv:0802.3318 [hep-th]];\\
  %%CITATION = PRLTA,100,191601;%%
 A.~Buchel and R.~C.~Myers,
  ``Causality of Holographic Hydrodynamics,''
  JHEP {\bf 0908}, 016 (2009)
  [arXiv:0906.2922 [hep-th]].
  %%CITATION = ARXIV:0906.2922;%%

\bibitem %[hofman]
 {hofman} D.~M.~Hofman,
``Higher Derivative Gravity, Causality and Positivity of Energy in a UV
complete QFT,''
  Nucl.\ Phys.\  B {\bf 823}, 174 (2009)
  [arXiv:0907.1625 [hep-th]].
  %%CITATION = NUPHA,B823,174;%%

\bibitem %[Sin]
 {Sin} X.~H.~Ge and S.~J.~Sin,
  ``Shear viscosity, instability and the upper bound of the Gauss-Bonnet
  coupling constant,''
  JHEP {\bf 0905}, 051 (2009)
  [arXiv:0903.2527 [hep-th]];\\
  %%CITATION = JHEPA,0905,051;%%
  R.~G.~Cai, Z.~Y.~Nie and Y.~W.~Sun,
  ``Shear Viscosity from Effective Couplings of Gravitons,''
  Phys.\ Rev.\  D {\bf 78}, 126007 (2008)
  [arXiv:0811.1665 [hep-th]];\\
  %%CITATION = PHRVA,D78,126007;%%
 R.~G.~Cai, Z.~Y.~Nie, N.~Ohta and Y.~W.~Sun,
``Shear Viscosity from Gauss-Bonnet Gravity with a Dilaton Coupling,''
  Phys.\ Rev.\  D {\bf 79}, 066004 (2009)
  [arXiv:0901.1421 [hep-th]];\\
  %%CITATION = PHRVA,D79,066004;%%
J.~de Boer, M.~Kulaxizi and A.~Parnachev,
  ``$AdS_7/CFT_6$, Gauss-Bonnet Gravity, and Viscosity Bound,''
  arXiv:0910.5347 [hep-th];\\
  %%CITATION = ARXIV:0910.5347;%%
X.~O.~Camanho and J.~D.~Edelstein, ``Causality constraints in AdS/CFT
from conformal collider physics and Gauss-Bonnet gravity,''
  arXiv:0911.3160 [hep-th];\\
  %%CITATION = ARXIV:0911.3160;%%
A.~Buchel, J.~Escobedo, R.~C.~Myers, M.~F.~Paulos, A.~Sinha and
M.~Smolkin, ``Holographic GB gravity in arbitrary dimensions,''
  arXiv:0911.4257 [hep-th].
  %%CITATION = ARXIV:0911.4257;%%

\bibitem %[jan3]
 {jan3} J.~de Boer, M.~Kulaxizi and A.~Parnachev,
 ``Holographic Lovelock Gravities and Black Holes,''
  arXiv:0912.1877 [hep-th];\\
  %%CITATION = ARXIV:0912.1877;%%
X.~O.~Camanho and J.~D.~Edelstein, ``Causality in AdS/CFT and Lovelock
theory,'' arXiv:0912.1944 [hep-th].
  %%CITATION = ARXIV:0912.1944;%%

\bibitem %[moreunpub]
 {moreunpub} Miguel Paulos, unpublished.

\bibitem %[qthydro]
 {qthydro} R.~C.~Myers, M.~F.~Paulos and A.~Sinha,
  ``Holographic studies of quasi-topological gravity,''
  arXiv:1004.2055 [hep-th].
  %%CITATION = ARXIV:1004.2055;%%


\bibitem %[aninda]
{aninda} A.~Sinha, ``On the new massive gravity and AdS/CFT,''
  arXiv:1003.0683 [hep-th].
  %%CITATION = ARXIV:1003.0683;%%

\bibitem %[newer]
{newer} J.~Oliva and S.~Ray, ``A new cubic theory of gravity in five
dimensions: Black hole, Birkhoff's theorem and C-function,''
  arXiv:1003.4773 [gr-qc].
  %%CITATION = ARXIV:1003.4773;%%

\bibitem %[newer2]
{newer2} J.~Oliva and S.~Ray, ``A Classification of Six Derivative
Lagrangians of Gravity and Static Spherically Symmetric Solutions,''
  arXiv:1004.0737 [gr-qc].
  %%CITATION = ARXIV:1004.0737;%%

\bibitem %[lovelock]
 {lovelock} D.~Lovelock,
  ``The Einstein tensor and its generalizations,''
  J.\ Math.\ Phys.\  {\bf 12}, 498 (1971);
  %%CITATION = JMAPA,12,498;%%
Aequationes Math. {\bf 4}, 127 (1970).

\bibitem %[GBheter]
 {GBheter} B.~Zwiebach,
  ``Curvature Squared Terms And String Theories,''
  Phys.\ Lett.\  B {\bf 156} (1985) 315.
  %%CITATION = PHLTA,B156,315;%%

\bibitem %[GBghost]
 {GBghost} D.~G.~Boulware and S.~Deser,
  ``String Generated Gravity Models,''
  Phys.\ Rev.\ Lett.\  {\bf 55} (1985) 2656.
  %%CITATION = PRLTA,55,2656;%%

\bibitem %[GBbh]
 {GBbh} J.~T.~Wheeler, ``Symmetric Solutions To The
    Gauss-Bonnet Extended Einstein Equations,''
  Nucl.\ Phys.\  B {\bf 268} (1986) 737;\\
  %%CITATION = NUPHA,B268,737;%%
J.~T.~Wheeler,
``Symmetric Solutions To The Maximally Gauss-Bonnet Extended Einstein
Equations,''
  Nucl.\ Phys.\  B {\bf 273} (1986) 732;\\
  %%CITATION = NUPHA,B273,732;%%
R.~C.~Myers and J.~Z.~Simon,
``Black Hole Thermodynamics in Lovelock Gravity,''
  Phys.\ Rev.\  D {\bf 38} (1988) 2434;\\
  %%CITATION = PHRVA,D38,2434;%%
R.~C.~Myers and J.~Z.~Simon,
  ``Black Hole Evaporation and Higher Derivative Gravity,''
  Gen.\ Rel.\ Grav.\  {\bf 21}, 761 (1989).
  %%CITATION = GRGVA,21,761;%%


\bibitem %[GBads]
{GBads} R.~G.~Cai,
``Gauss-Bonnet black holes in AdS spaces,''
  Phys.\ Rev.\  D {\bf 65} (2002) 084014
  [arXiv:hep-th/0109133].
  %%CITATION = PHRVA,D65,084014;%%

\bibitem %[GBads2]
 {GBads2} S.~Nojiri and S.~D.~Odintsov, ``Anti-de
    Sitter black hole thermodynamics in higher derivative gravity and
    new confining-deconfining phases in dual CFT,''
  Phys.\ Lett.\  B {\bf 521} (2001) 87
  [Erratum-ibid.\  B {\bf 542} (2002) 301]
  [arXiv:hep-th/0109122];\\
  %%CITATION = PHLTA,B521,87;%%
  Y.~M.~Cho and I.~P.~Neupane,
  ``Anti-de Sitter black holes, thermal phase transition and holography in
  higher curvature gravity,''
  Phys.\ Rev.\  D {\bf 66}  (2002) 024044
  [arXiv:hep-th/0202140];\\
  %%CITATION = PHRVA,D66,024044;%%
  I.~P.~Neupane,
  ``Black hole entropy in string-generated gravity models,''
  Phys.\ Rev.\  D {\bf 67} (2003) 061501
  [arXiv:hep-th/0212092];\\
  %%CITATION = PHRVA,D67,061501;%%
  I.~P.~Neupane,
  ``Thermodynamic and gravitational instability on hyperbolic spaces,''
  Phys.\ Rev.\  D {\bf 69} (2004) 084011
  [arXiv:hep-th/0302132].
  %%CITATION = PHRVA,D69,084011;%%

\bibitem %[genGibHawk]
 {genGibHawk} R.~C.~Myers, ``Higher Derivative Gravity,
Surface Terms and String Theory,''
  Phys.\ Rev.\  D {\bf 36}, 392 (1987).
  %%CITATION = PHRVA,D36,392;%%

\bibitem %[counter]
 {counter} V.~Balasubramanian and P.~Kraus,
  ``A stress tensor for anti-de Sitter gravity,''
  Commun.\ Math.\ Phys.\  {\bf 208}, 413 (1999)
  [arXiv:hep-th/9902121];\\
  %%CITATION = CMPHA,208,413;%%
R.~Emparan, C.~V.~Johnson and R.~C.~Myers,
  ``Surface terms as counterterms in the AdS/CFT correspondence,''
  Phys.\ Rev.\  D {\bf 60}, 104001 (1999)
  [arXiv:hep-th/9903238];\\
  %%CITATION = PHRVA,D60,104001;%%
R.~B.~Mann,
  ``Misner string entropy,''
  Phys.\ Rev.\  D {\bf 60}, 104047 (1999)
  [arXiv:hep-th/9903229].
  %%CITATION = PHRVA,D60,104047;%%

\bibitem %[WaldEnt]
 {WaldEnt} R.~M.~Wald, ``Black hole entropy is the
    Noether charge,''
  Phys.\ Rev.\  D {\bf 48}, 3427 (1993)
  [arXiv:gr-qc/9307038];\\
  %%CITATION = PHRVA,D48,3427;%%
  V.~Iyer and R.~M.~Wald,
  ``Some properties of Noether charge and a proposal for dynamical black hole
  entropy,''
  Phys.\ Rev.\  D {\bf 50}, 846 (1994)
  [arXiv:gr-qc/9403028];\\
  %%CITATION = PHRVA,D50,846;%%
T.~Jacobson, G.~Kang and R.~C.~Myers,
  ``On Black Hole Entropy,''
  Phys.\ Rev.\  D {\bf 49}, 6587 (1994)
  [arXiv:gr-qc/9312023].
  %%CITATION = PHRVA,D49,6587;%%


\bibitem %[topo]
 {topo} T.~Eguchi, P.~B.~Gilkey and A.~J.~Hanson,
``Gravitation, Gauge Theories And Differential Geometry,''
  Phys.\ Rept.\  {\bf 66}, 213 (1980).
  %%CITATION = PRPLC,66,213;%%

\bibitem %[AdSCFT]
 {AdSCFT} O.~Aharony, S.~S.~Gubser, J.~M.~Maldacena,
    H.~Ooguri and Y.~Oz,
  ``Large N field theories, string theory and gravity,''
  Phys.\ Rept.\  {\bf 323}, 183 (2000)
  [arXiv:hep-th/9905111].
  %%CITATION = PRPLC,323,183;%%

\bibitem %[hawkpage]
 {hawkpage} S.~W.~Hawking and D.~N.~Page,
  ``Thermodynamics Of Black Holes In Anti-De Sitter Space,''
  Commun.\ Math.\ Phys.\  {\bf 87}, 577 (1983);\\
  %%CITATION = CMPHA,87,577;%%
E.~Witten, ``Anti-de Sitter space, thermal phase transition, and
confinement in  gauge theories,''
  Adv.\ Theor.\ Math.\ Phys.\  {\bf 2}, 505 (1998)
  [arXiv:hep-th/9803131].
  %%CITATION = 00203,2,505;%%

\bibitem %[Roberto]
 {Roberto} R.~Emparan, ``AdS/CFT duals of topological black holes and
the entropy of zero-energy states,''
  JHEP {\bf 9906}, 036 (1999)
  [arXiv:hep-th/9906040].
  %%CITATION = JHEPA,9906,036;%%

\bibitem %[Mann]
 {Mann} M.~H.~Dehghani and R.~B.~Mann,
  ``Thermodynamics of rotating charged black branes in third order Lovelock
  gravity and the counterterm method,''
  Phys.\ Rev.\  D {\bf 73}, 104003 (2006)
  [arXiv:hep-th/0602243];\\
  %%CITATION = PHRVA,D73,104003;%%
M.~H.~Dehghani and R.~B.~Mann, ``Thermodynamics of rotating charged
black branes in third order Lovelock gravity and the counterterm
method,''
  Phys.\ Rev.\  D {\bf 73}, 104003 (2006)
  [arXiv:hep-th/0602243];\\
  %%CITATION = PHRVA,D73,104003;%%
M.~H.~Dehghani and R.~Pourhasan,
  ``Thermodynamic instability of black holes of third order Lovelock gravity,''
  Phys.\ Rev.\  D {\bf 79}, 064015 (2009)
  [arXiv:0903.4260 [gr-qc]];\\
  %%CITATION = PHRVA,D79,064015;%%
M.~H.~Dehghani and M.~Shamirzaie,
  ``Thermodynamics of asymptotic flat charged black holes in third order
  Lovelock gravity,''
  Phys.\ Rev.\  D {\bf 72}, 124015 (2005)
  [arXiv:hep-th/0506227].\\
  S.~H.~Hendi and M.~H.~Dehghani,
  ``Taub-NUT Black Holes in Third order Lovelock Gravity,''
  Phys.\ Lett.\  B {\bf 666}, 116 (2008)
  [arXiv:0802.1813 [hep-th]].

\bibitem %[EntMatch]
 {EntMatch} V.~Iyer and R.~M.~Wald,
  ``A Comparison of Noether charge and Euclidean methods for computing the
  entropy of stationary black holes,''
  Phys.\ Rev.\  D {\bf 52}, 4430 (1995)
  [arXiv:gr-qc/9503052].
  %%CITATION = PHRVA,D52,4430;%%

\bibitem %[Cubevar]
 {Cubevar} Y.~D\'ecanini and A.~Folacci
 ``Irreducible Forms for the Metric Variations of the Action Terms
 of Sixth-Order Gravity and Approximated Stress-Energy Tensor,''
  Class.\ Quant.\ Grav.\ {\bf 24} (2007) 4777
  [arXiv:hep-th/0706.0691]
  %%CITATION = CQRGD,24,4777;%%

\bibitem %[standard]
 {standard} For example, see:\\
H.~Liu and A.~A.~Tseytlin,
  ``D = 4 super Yang-Mills, D = 5 gauged supergravity, and D = 4 conformal
  supergravity,''
  Nucl.\ Phys.\  B {\bf 533}, 88 (1998)
  [arXiv:hep-th/9804083];\\
  %%CITATION = NUPHA,B533,88;%%
G.~Arutyunov and S.~Frolov,
  ``Three-point Green function of the stress-energy tensor in the AdS/CFT
  correspondence,''
  Phys.\ Rev.\  D {\bf 60}, 026004 (1999)
  [arXiv:hep-th/9901121].
  %%CITATION = PHRVA,D60,026004;%%

\bibitem %[ozzy]
 {ozzy} H.~Osborn and A.~C.~Petkou,
``Implications of Conformal Invariance in Field Theories for General
  Dimensions,''
  Annals Phys.\  {\bf 231}, 311 (1994)
  [arXiv:hep-th/9307010];\\
  %%CITATION = APNYA,231,311;%%
J.~Erdmenger and H.~Osborn, ``Conserved currents and the
energy-momentum tensor in conformally  invariant theories for general
dimensions,''
  Nucl.\ Phys.\  B {\bf 483}, 431 (1997)
  [arXiv:hep-th/9605009].
  %%CITATION = NUPHA,B483,431;%%

\bibitem %[HM]
 {HM} D.~M.~Hofman and J.~Maldacena,
  ``Conformal collider physics: Energy and charge correlations,''
  JHEP {\bf 0805}, 012 (2008)
  [arXiv:0803.1467 [hep-th]].
  %%CITATION = JHEPA,0805,012;%%

\bibitem %[ctheorem]
{ctheorem} R.~C.~Myers and A.~Sinha, ``Seeing a c-theorem with
holography,''
  arXiv:1006.1263 [hep-th].
  %%CITATION = ARXIV:1006.1263;%%

\end{thebibliography}
\end{document}